\documentclass[a4paper,twocolumn,10pt,accepted=2021-03-09]{quantumarticle}
\pdfoutput=1
\usepackage[utf8]{inputenc}
\usepackage[english]{babel}
\usepackage[T1]{fontenc}
\usepackage{amsmath,amssymb}
\usepackage{hyperref}
\usepackage{tabularx}
\usepackage[numbers,sort&compress]{natbib}

\usepackage{tikz}
\usepackage{lipsum}

\def\bra#1{\langle #1|}
\def\ket#1{|#1 \rangle}
\def\bracket#1#2{\langle #1|#2 \rangle}

\begin{document}

\title{Complete Information Balance in Quantum Measurement}

\author{Seung-Woo Lee}
\email{swleego@gmail.com}
\affiliation{Center for Quantum Information, Korea Institute of Science and Technology, Seoul, 02792, Korea}
\affiliation{Quantum Universe Center, Korea Institute for Advanced Study, Seoul 02455, Korea}
\orcid{0000-0003-1006-0333}
\author{Jaewan Kim}
\affiliation{School of Computational Sciences, Korea Institute for Advanced Study, Seoul 02455, Korea}
\author{Hyunchul Nha}
\affiliation{Department of Physics, Texas A\&M University at Qatar, Education City, P.O.Box 23874, Doha, Qatar}
\maketitle

\begin{abstract}
Quantum measurement is a basic tool to manifest intrinsic quantum effects from fundamental tests to quantum information applications. 
While a measurement is typically performed to gain information on a quantum state, its role in quantum technology is indeed manifold. For instance, quantum measurement is a crucial process element in measurement-based quantum computation. It is also used to detect and correct errors thereby protecting quantum information in error-correcting frameworks.
It is therefore important to fully characterize the roles of quantum measurement encompassing information gain, state disturbance and reversibility, together with their fundamental relations. 
Numerous efforts have been made to obtain the trade-off between information gain and state disturbance, which becomes a practical basis for secure information processing.
However, a complete information balance is necessary to include the reversibility of quantum measurement, which constitutes an integral part of practical quantum information processing.
We here establish all pairs of trade-off relations involving information gain, disturbance, and reversibility, and crucially the one among all of them together. By doing so, we show that the reversibility plays a vital role in completing the information balance. 
Remarkably, our result can be interpreted as an information-conservation law of quantum measurement in a nontrivial form. We completely identify the conditions for optimal measurements that satisfy the conservation for each tradeoff relation with their potential applications. Our work can provide a useful guideline for designing a quantum measurement in accordance with the aims of quantum information processors.
\end{abstract}

%%%%%%%%%%%%%Introduction
\section{Introduction}

Since the early days of quantum mechanics, quantum measurement has been one of the central topics in quantum theory \cite{Heisenberg}. Unlike the measurement in classical world that only passively provides information on a physical state, quantum measurement has a critical role to reveal quantum phenomena from fundamental tests to quantum technology applications. For instance, incompatible quantum measurements are essential to demonstrate quantum nonlocality \cite{QN1,QN2}.  
Quantum measurement is also a crucial tool in quantum information protocols for not only extracting information but also processing computation in one-way quantum computation \cite{OQ}, protecting quantum information by detecting and correcting errors in quantum error-correcting protocols, {\it etc}. \cite{QIQM,Wiseman09,Kurt,teleportation,MBQC,AdvancedTele,AdvancedMetro,QEC}. That is, the role of quantum measurement is manifold in quantum technology, which requires a thorough characterization encompassing all relevant aspects of measurement.

A well-known effect of quantum measurement is its inevitable disturbance on quantum states, which constitutes a profound basis for secure quantum information processing, e.g. quantum cryptography \cite{Gisin2002,BB84}. 
It has been a topic of fundamental and practical importance to identify information balance in quantum measurement, particularly the trade-off relation between information gain and state disturbance \cite{Groenewold71,Lindblad72,Ozawa86,Fuchs96,Fuchs01,Banaszek01,Banaszek02,Dariano03,Sacchi06,Buscemi08,Luo10,Berta14}.
On the other hand, the usual notion that quantum measurement is irreversible has been substantially reassessed in recent years \cite{Ueda92,Royer95,Ueda96,Jordan10,Terashima11,Terashima16,Cheong12,Koashi99,Terashima03,Korotkov06,YSKim09,Korotkov10,YSKim11,Katz08,Schindler13,JCLee11,HTLim14,Chen14}. 
It turns out that a quantum measurement weakly interacting with a system can be reversed even to faithfully recover the input state with a non-zero success probability \cite{Ueda92,Royer95}. 
This issue of reversibility thus becomes an integral part to consider in characterizing a quantum measurement in its entirety \cite{Ueda92,Royer95,Ueda96,Jordan10}. 
The measurement reversibility has also widely applied to range of applications, e.g.~quantum error corrections \cite{Koashi99}, gate operations \cite{Terashima03,Korotkov06} and decoherence suppression \cite{YSKim09,Korotkov10,YSKim11}. 
Reversing the disturbance of quantum measurement was experimentally demonstrated with superconducting \cite{Katz08}, trapped-ion \cite{Schindler13} and photonic qubits \cite{YSKim09,YSKim11,JCLee11}. 
Along this line, a trade-off of the reversibility against information gain in quantum measurement was rigorously assessed \cite{Cheong12,HTLim14,Chen14}. Trade-offs among information gain, state disturbance and reversibility were also quantitatively studied at the level of a single measurement outcome \cite{Terashima11,Terashima16}.

However, there does not exist a unified framework yet to deal with all the information contents together as universal quantities in general quantum measurement \cite{Berta14},  
which is essential to draw a complete picture on information balance in quantum measurement.
This is very important for a broad applicability, e.g. designing a quantum measurement to be adapted to the aim of quantum information processing at hand.
%Therefore, in order to complete the total information balance in quantum measurement, verifying full quantitative relations between all the obtained, disturbed, and reversible information is essential but has been missing thus far.
Here we establish such a complete information balance in quantum measurement. We derive the full trade-off relations among those three information contents, ~information gain $\cal G$, disturbance $\cal D$, and reversibility $\cal R$. %which are defined to be universal (i.e.,~independent of the input quantum state) with a clear operational meaning.
We particularly show that the reversibility $\cal R$ plays a crucial role in information balance accounting for the gap between $\cal G$ and $\cal D$. That is, we find that the global trade-off relation ($\cal G$-$\cal D$-$\cal R$) tightens the trade-off between the gain and disturbance ($\cal G$-$\cal D$) only \cite{Banaszek01}. On the other hand, the trade-off between disturbance and reversibility ($\cal D$-$\cal R$) compensates the gain and reversibility relation ($\cal G$-$\cal R$). Consequently, our result shows that the total information is balanced in an ideal quantum measurement process, which can be interpreted as a conservation of information contents in quantum measurement. 

We fully obtain the conditions to saturate all of the trade-off relations and thereby define an {\it optimal} quantum measurement providing maximal information contents quantum mechanics fundamentally allows. 
Our framework can offer useful guidelines for designing measurement-based quantum information protocols in quantum computation \cite{MBQC}, teleportation \cite{AdvancedTele}, quantum metrology \cite{AdvancedMetro}, and quantum error corrections \cite{QEC}, {\it etc}.. As all the information contents defined here are directly measurable, our results are readily testable and applicable to any quantum information platforms. Our work may contribute to deepening our fundamental understanding of quantum measurement and provide a rigorous practical benchmark to optimize measurement-based quantum information protocols.

%%%%%%%%%%%%%%%

\section{General framework}

\begin{figure}[t]
\centering
\includegraphics[width=1.0\linewidth]{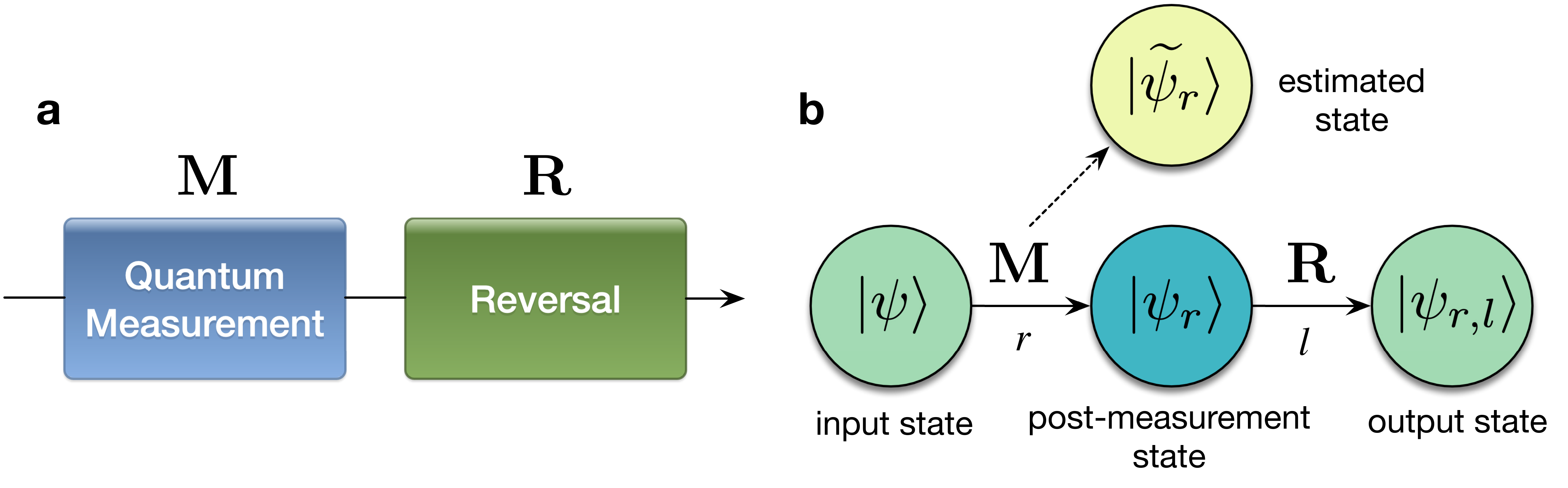}
\caption{(a) Quantum measurement ${\bf M}$ and reversal process $\bf R$. (b) The input, the post-measurement, the estimated and the output state in a single trial, when the outcomes of the measurement and the reversal are $r$ and $l$, respectively, are denoted by $\ket{\psi}$, $\ket{\psi_r}$, $\ket{\widetilde{\psi}_r}$, and $\ket{\psi_{r,l}}$, respectively. The closeness of those states on average determines the information contents of a given quantum measurement ${\bf M}$.
}\label{fig:fig1}
\end{figure}

We begin with the general framework for addressing the information changes by a quantum measurement $\bf M$. Assume that an arbitrary quantum state $\rho$ is prepared to convey information. We perform a quantum measurement $\bf M$ to extract the information, which disturbs and changes the input state $\rho$ to another state. We then apply a subsequent reversing operation $\bf R$ to characterize the reversibility of $\bf M$ (see Fig.~\ref{fig:fig1}(a)). The reversing operation is assumed to be chosen to recover the input state $\rho$, i.e., $({\bf R}\circ{\bf M})(\rho) \propto \rho$.

Without loss of generality, we assume here that the input information is encoded onto pure states $\rho=\ket{\psi}\bra{\psi}$ in $d$-dimensional Hilbert space, but the results derived in what follows are valid for any mixed input states. Consider a quantum measurement described by a set of operators ${\bf M}=\{\hat{M}_r|r=1,\ldots, n\}$, satisfying the completeness relation $\sum_r\hat{M}^{\dag}_r\hat{M}_r=\hat{\openone}$ \cite{Wiseman09,Kurt}. When the measurement outcome is $r$, the input state is changed to $\ket{\psi_r}=\hat{M}_r\ket{\psi}/{\sqrt{p(r,\psi)}}$ where $p(r,\psi)=\bra{\psi}\hat{M}_r^{\dagger}\hat{M}_r\ket{\psi}$. For each outcome $r$, we may estimate the input state as $\ket{\widetilde{\psi}_r}$ by a certain estimation strategy. We then apply a subsequent reversing operation described by a set of operators ${\bf R}=\{\hat{R}_{r,l}|l=1,\ldots ,m \}$, satisfying the completeness relation $\sum_l\hat{R}^{\dag}_{r,l}\hat{R}_{r,l}=\hat{\openone}$ for each $r$. The final output state is then given by $\ket{\psi_{r,l}}=\hat{R}_{r,l}\hat{M}_r\ket{\psi}/{\sqrt{p(r,l,\psi)}}$, where $p(r,l,\psi)=\bra{\psi}\hat{M}_r^{\dag}\hat{R}^\dag_{r,l}\hat{R}_{r,l}\hat{M}_r\ket{\psi}$. The reversing operation $\bf R$ here is assumed to be appropriately chosen according to the measurement outcome $r$. We  note that the whole process cannot be described by a unitary operation as it can be probabilistic and conditional on the result of $\bf M$ and observer's choice of $\bf R$. Note that our framework differs from the Petz recovery map \cite{Petz86,Petz88}. 

%%%%%%%%%%%%%%%%%%%%%%%%%%%%%%%%%%%%%%%
%%%%%%%%%%%%%%%%%%%%%%%%%%%%%%%%%%%%%%%%%

%%%%%%%%%%%%%%%%%

\section{Information contents}

Let us define the information contents within the above general framework. For a given quantum measurement, the changes of information due to the measurement $\bf M$ and the reversing operation $\bf R$ can be defined in terms of the closeness between the input $\ket{\psi}$, the post-measurement $\ket{\psi_r}$, the estimated $\ket{\widetilde{\psi}_r}$, and the output $\ket{\psi_{r,l}}$ states. See Fig.~\ref{fig:fig1}(b) together with the details of the definitions in Appendix~\ref{asec:Def}.

(i) {\it Information Gain}: The amount of information obtained by $\bf M$ can be quantified based on the overlap between the input $\ket{\psi}$ and the estimated state $\ket{\widetilde{\psi}_r}$. The estimation fidelity is obtained by averaging $|\bracket{\widetilde{\psi}_r}{\psi}|^2$ over all input states and possible measurement outcomes, i.e., $ \int d \psi \sum_{r=1}^{n} p(r,\psi) |\langle \widetilde{\psi}_r | \psi \rangle|^2$, which has different values depending on the estimation strategies. We define the {\em information gain} as the maximum estimation fidelity over all possible  strategies, which can be obtained as (see Appendix~\ref{asec:IG}),
\begin{equation}
\label{eq:estifide} {\cal G}=\frac{1}{d(d+1)}\bigg[d+\sum_{r=1}^n (\lambda_0^r)^2\bigg],
\end{equation}
where $\lambda_0^r$ is the largest singular value of $\hat{M}_r$ in the singular value decomposition (Appendix~\ref{asec:SVD}) \cite{Banaszek01}. The information gain thus lies in the range $1/d \leq {\cal G} \leq 2/(d+1)$, where the upper bound is reached by a von Neumann measurement and the lower bound by a unitary operation or a random guess.

(ii) {\it Disturbance}: In order to quantify the amount of disturbance by $\bf M$, we consider the {\em operation fidelity} first. The operation fidelity of $\bf M$ is given by averaging the overlap between the input $\ket{\psi}$ and the post-measurement state $\ket{\psi_r}$, i.e., $\int d \psi \sum_{r=1}^{n} p(r,\psi) |\langle \psi_r | \psi \rangle|^2$. We can then evaluate the maximum of the operation fidelity (see Appendix~\ref{asec:MaxF}), resulting in 
\begin{equation}
\label{eq:maxopfid} {\cal F}=\frac{1}{d(d+1)}\bigg[d+\sum_{r=1}^n \Big(\sum_{i=0}^{d-1}\lambda_i^r\Big)^2\bigg].
\end{equation}
In this case, all the singular values $\lambda_i^r$ of $\hat{M}_r$ are involved in determining the average fidelity of the disturbed states. We may then define the {\em disturbance} induced by $\bf M$ as the minimum operation infidelity given by
\begin{equation}
{\cal D} =1-{\cal F},
\end{equation}
scaled in the range $0 \leq {\cal D} \leq (d-1)/(d+1)$.

Similarly, the operation fidelity of the overall process ${\bf R}\circ{\bf M}$ without postselection can be obtained by evaluating the average fidelity between the input $\ket{\psi}$ and output state $\ket{\psi_{r,l}}$, i.e., $\int d\psi \sum^n_{r=1} \sum^m_{l=1}P(r,l,\psi) |\langle \psi_{r,l} | \psi \rangle|^2$. Its maximum can be obtained as (see Appendix~\ref{asec:MaxgRev})
\begin{equation}
\label{eq:maxgenrev} 
{\cal F}({\bf R}\circ{\bf M})=\frac{1}{d(d+1)}\bigg[d+\sum_{r=1}^n  \sum^m_{l=1} \Big(\sum_{i=0}^{d-1}\lambda_i^r\lambda_i^{r,l}\Big)^2\bigg],
\end{equation}
where the singular values $\lambda_i^{r,l}$ of $\hat{R}_{r,l}$ are additionally involved in determining the output fidelity. The operation fidelity lies in the range $2/(d+1) \leq {\cal F} \leq 1$, where the upper bound is reached by a unitary operation and the lower bound by a von Neumann measurement. 

(iii) {\it Reversibility}: We consider the success event of the reversing operation ${\bf R}$, which faithfully recovers the input state, i.e., $\ket{\psi_{r,l}}\propto\ket{\psi}$, to evaluate the reversibility. Assume that the operators $\hat{R}_{r,l}$ for $l=1,\ldots,s<m$ are associated with the success events s.t. $\hat{R}_{r,l}\hat{M}_r\ket{\psi}=\eta_{r,l}\ket{\psi}$, where $|\eta_{r,l}|^2$ is the success probability of the reversing operation when the outcome is $r$. The {\em reversibility} is then obtained as the maximum overall success probability,
\begin{equation}
\label{eq:revmax}
{\cal R}=\max_{\{\hat{R}_{r,l}\}} \int d\psi \sum^n_{r=1}\sum^{s<m}_{l=1}| \bra{\psi}\hat{R}_{r,l}\hat{M}_r\ket{\psi}|^2=\sum_{r=1}^{n}(\lambda^r_{d-1})^2,
\end{equation}
where $\lambda^r_{d-1}$ is the smallest singular value of $\hat{M}_r$ \cite{Cheong12}. The reversibility is scaled as $0\leq {\cal R}\leq 1$. 

The optimal reversing operator can be defined in the following context. Assume that the measurement operator $\hat{M}_r$ is represented in the singular value decomposition as $\hat{M}_r=\hat{V}_r\hat{D}_r$ with a unitary operator $\hat{V}_r$ and a diagonal matrix $\hat{D}_r=\sum_{i=0}^{d-1}\lambda^r_i\ket{i} \bra{i}$ (see Appendix~\ref{asec:SVD}). Its optimal reversing operator can then be written by $\hat{R}_{r,1}=\lambda^r_{d-1}\hat{D}^{-1}_r\hat{V}^{\dag}_r$, where $\hat{D}^{-1}_r=\sum_{i=0}^{d-1} (\lambda^r_i)^{-1} \ket{i}\bra{i}$ with nonzero $\lambda^r_i$ (Here, we set $s=1$ without loss of generality). For example, a quantum measurement described by $\hat{M}_1=\sqrt{\eta}\ket{1}\bra{1}$ and $\hat{M}_2=\ket{0}\bra{0}+\sqrt{1-\eta}\ket{1}\bra{1}$ with measurement strength $0\leq \eta \leq1$, can be optimally reversed by $\hat{R}_{2,1}=\sqrt{1-\eta}\ket{0}\bra{0}+\ket{1}\bra{1}$ and $\hat{R}_{2,2}=\sqrt{\eta}\ket{0}\bra{0}$ for the measurement outcome $r=2$ and the reversal process $l=1$, respectively. In this case, the reversibility is given by ${\cal R}=1-\eta$ from Eq.~\eqref{eq:revmax}. Note that a unitary operation (no measurement) is deterministically reversible ${\cal R}=1$ while a von Neumann measurement is completely irreversible ${\cal R}=0$. 

We now have three information contents characterizing a quantum measurement, i.e.,~information gain ${\cal G}$, disturbance ${\cal D}$, and reversibility ${\cal R}$. These are universal quantities averaged over all input states of a given dimension $d$ and have a clear operational meaning in terms of quantum fidelity, fulfilling the requirements for the information contents in quantum measurement \cite{Berta14}. Note that the information contents evaluated here with pure input states~$\rho=\ket{\psi}\bra{\psi}$ and the accompanying consequences in information balance are generally valid for arbitrary mixed input states $\rho$, since the maximum averaged in the space of pure states must represent the maximum in the space of convex combination of pure states, i.e.~mixed states.\\

%%%%%%%%%%%%%%%%
%\section{Trade-off relations}

\section{Information balance in quantum measurement}

In this section, we now derive the trade-off relations among the information contents of quantum measurement. Before presenting those trade-off relations, we first introduce and prove a useful inequality on the relation between the reversibility $\cal R$ and the overall operation fidelity ${\cal F}({\bf R}\circ{\bf M})$ after the measurement followed by the reversing operation defined in the previous section:\\

\noindent
{\bf Lemma 1}. For any quantum measurement ${\bf M}$ and subsequent reversing operation ${\bf R}$, the reversibility ${\cal R}$ and the operation fidelity of the overall process ${\cal F}({\bf R}\circ{\bf M})$ always satisfy an inequality 
\begin{equation}
\label{eq:lemma1}
2+(d-1){\cal R} \leq (d+1){\cal F}({\bf R}\circ{\bf M}),
\end{equation}
in arbitrary finite dimension $d\geq2$.\\

Detailed proof of Lemma 1 is given in Appendix~\ref{asec:Lemma1}. We define a vector $\vec{u}^l_i=(\lambda^{r=1}_i\lambda^{r=1,l}_i, \ldots , \lambda^{r=n}_i \lambda^{r=n,l}_i)$ for $i=0,\ldots,d-1$ in terms of the singular values $\lambda^{r}_i$ of the measurement operators $\{\hat{M}_r\}$ and $\lambda^{r,l}_i$ of the reversing operators $\{\hat{R}_{r,l}\}$. The equality in \eqref{eq:lemma1} holds if and only if the quantum measurement satisfies 
\begin{equation}
\vec{u}^l_i\cdot \vec{u}^l_j=\delta_{ij}|\vec{u}^l_i|^2,~\forall l\neq1.
\end{equation}
We will use Lemma 1 to derive the trade-off relations presented in what follows.

Before presenting our main results, we introduce two  previously known trade-off relations between information contents, denoted by $\cal G$-$\cal D$ \cite{Banaszek01} and $\cal G$-$\cal R$ \cite{Cheong12} as below.

($\cal G$-$\cal D$: Information gain and Disturbance trade-off) The trade-off relation between the information gain ${\cal G}$ and disturbance ${\cal D}=1-F$ was derived in Ref.~\cite{Banaszek01} as
\begin{equation}
\label{eq:banaszek}
\sqrt{{\cal F}-\frac{1}{d+1}}\leq \sqrt{{\cal G}-\frac{1}{d+1}} +\sqrt{(d-1)\bigg(\frac{2}{d+1}-{\cal G}\bigg)}.
\end{equation}
This is the quantitative proof of the heuristic knowledge `the more information a quantum measurement obtains, the more disturbed the quantum state becomes.' 
It gives the lower bound of diturbance for a given amount of information that a quantum measurement extracts.

($\cal G$-$\cal R$: Information gain and Reversibility trade-off) The trade-off relation between the information gain $\cal G$ and the reversibility $\cal R$ was derived in Ref.~\cite{Cheong12} as
\begin{equation}
\label{eq:balance} d(d+1){\cal G}+(d-1){\cal R} \leq 2d,
\end{equation}
which was the first information-theoretic approach introducing the role of the reversibility in quantum measurement. 
%It captures the idea that `the more information a quantum measurement obtains, the less reversible the quantum state becomes.' 
It captures the idea that `the more information a quantum measurement obtains, the less reversible the quantum measurement is.' 

\begin{figure}
\centering
\includegraphics[width=0.8\linewidth]{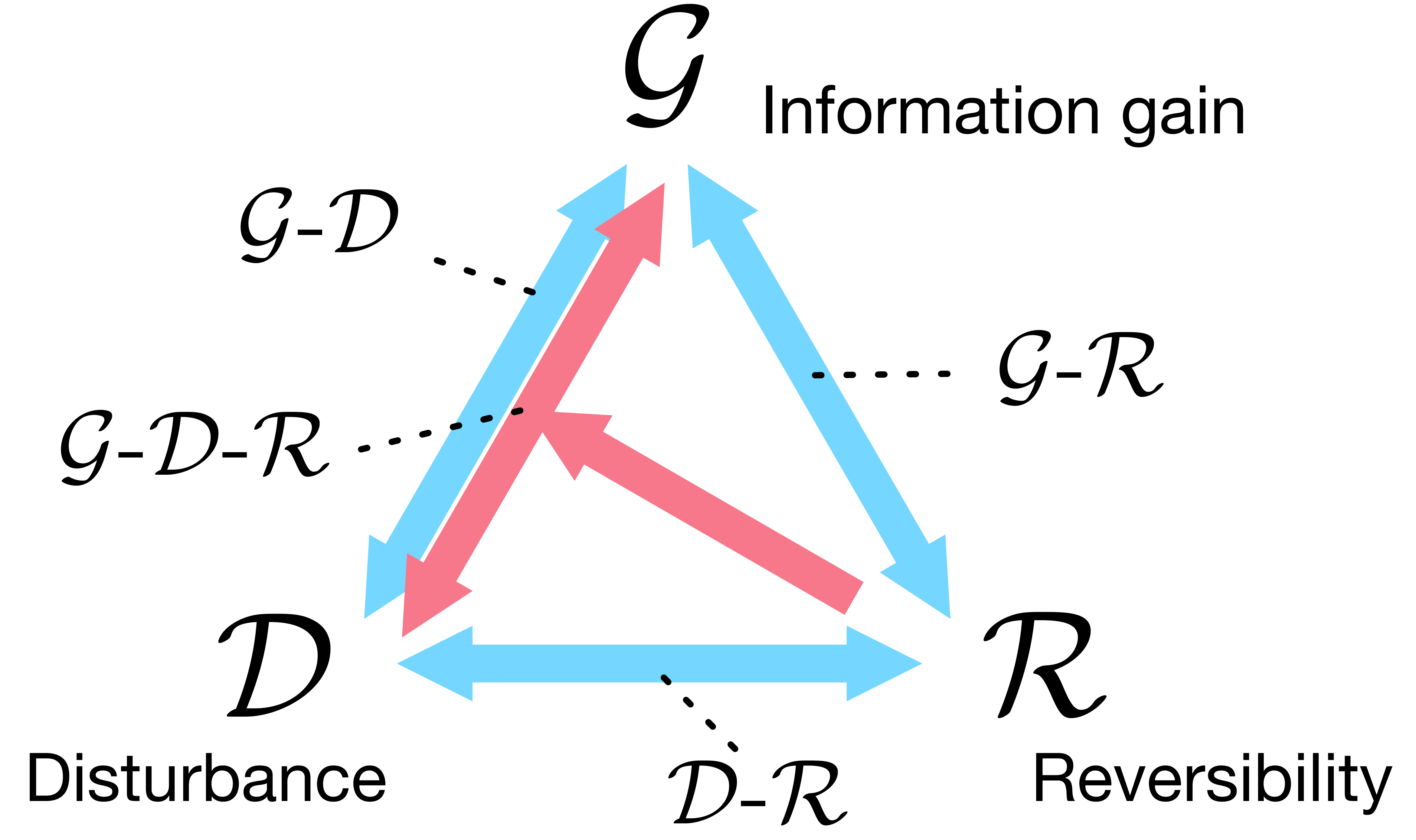}
\caption{Three information contents of quantum measurement, i.e., information gain ${\cal G}$, disturbance ${\cal D}=1-{\cal F}$, and reversibility ${\cal R}$, and their quantitative links: ($\cal G$-$\cal D$) in Eq.~\eqref{eq:banaszek} \cite{Banaszek01}, ($\cal G$-$\cal R$) in Eq.~\eqref{eq:balance} \cite{Cheong12}, ($\cal G$-$\cal D$-$\cal R$) Theorem 1 in Eq.~\eqref{eq:t1}, ($\cal D$-$\cal R$) Theorem 2 in Eq.~\eqref{eq:t2}.
}\label{fig:fig2}
\end{figure}

However, a global trade-off relation including all three information contents has been missing so far. 
The full quantitative links among the three information contents have also not been completed (see Fig.~\ref{fig:fig2}).
Let us now derive this global trade-off relation, including all the information contents, $\cal G$, $\cal D$, and $\cal R$, aiming to complete the total information balance as follows. \\

\noindent
{\bf Theorem 1}. ($\cal G$-$\cal D$-$\cal R$: Global trade-off relation) The information gain ${\cal G}$, the disturbance ${\cal D}$ and the reversibility ${\cal R}$ of quantum measurement always satisfy an inequality
\begin{align}
\label{eq:t1}
\sqrt{{\cal F}-\frac{1}{d+1}}&\leq \sqrt{{\cal G}-\frac{1}{d+1}} + \sqrt{\frac{\cal R}{d(d+1)}} \\
\nonumber
&\hspace{3mm}+\sqrt{(d-2)\bigg(\frac{2}{d+1}-{\cal G}-\frac{\cal R}{d(d+1)}\bigg)},
\end{align}
with ${\cal F}=1-{\cal D}$, in an arbitrary dimension $d\geq2$.\\

The inequality in \eqref{eq:t1} determines the quantitative relation among the three information contents, $\cal G$, $\cal D$, and $\cal R$ (see Appendix~\ref{asec:Theorem1} for the details of the proof). This relation can be interpreted in various ways. It draws the upper bound of $\cal F$ or equivalently the lower bound of $\cal D$ with respect to both ${\cal G}$ and ${\cal R}$. Or, it indicates the maximum possible reversibility for a given pair of the information amounts ${\cal G}$ and $\cal D$. 

When $d=2$, the inequality in \eqref{eq:t1} is equivalent to $\cal G$-$\cal D$ in \eqref{eq:banaszek}. This is due to the relation between the information gain ${\cal G}$ and the reversibility ${\cal R}$ for $d=2$, i.e. 
$\frac{2}{3}-{\cal G}= \frac{{\cal R}}{6}$. It is given by the completeness condition $\sum_r\hat{M}^{\dag}_r\hat{M}_r=\hat{\openone}$ leading to $\sum_{r=1}^{n}(\lambda^r_{0})^2+\sum_{r=1}^{n}(\lambda^r_{1})^2=d=2$ 
together with Eqs.~\eqref{eq:estifide} and ~\eqref{eq:revmax}.   

On the other hand, for $d>2$ in general, the $\cal G$-$\cal D$-$\cal R$ inequality in \eqref{eq:t1} is fundamentally different from $\cal G$-$\cal D$ in \eqref{eq:banaszek} and $\cal G$-$\cal R$ in \eqref{eq:balance}. 
Note that $\cal G$-$\cal D$-$\cal R$ provides a tighter bound than $\cal G$-$\cal D$ in the relation between the information gain and the disturbance. This becomes clearer with its saturation condition described below.\\

\noindent
($\cal G$-$\cal D$-$\cal R$ saturation condition) The $\cal G$-$\cal D$-$\cal R$ inequality \eqref{eq:t1} is saturated if and only if the quantum measurement satisfies following conditions: all $\vec{v}_i$ for $i=0,\cdots,d-1$ are collinear and 
\begin{equation}
\label{eq:scon}
|\vec{v}_1|=\cdots=|\vec{v}_{d-2}|,
\end{equation}
where $\vec{v}_i=(\lambda^{r=1}_i, \ldots , \lambda^{r=n}_i)$ is a vector defined with the singular values of $\hat{M}_r$.\\

See Appendix~\ref{asec:Theorem1} for the detailed description of the condition. We can see that a quantum measurement satisfying the $\cal G$-$\cal D$ saturation condition (i.e., all $\vec{v}_i$ are collinear and $|\vec{v}_1|=\cdots=|\vec{v}_{d-1}|$  \cite{Banaszek01}) also satisfies the saturation condition of $\cal G$-$\cal D$-$\cal R$ (Table~\ref{tab:table1}), but the converse is not always true. It indicates that the information balance can be more tightly characterized by $\cal G$-$\cal D$-$\cal R$ in a broader set of quantum measurements than $\cal G$-$\cal D$. In section~\ref{sec:Class}, we will further discuss on the different sets of quantum measurements classified based on the saturation conditions of the trade-off relations. Besides, we also analytically show that the right-hand side of the inequality \eqref{eq:t1} is always lower than or equal to the right-hand side of the inequality \eqref{eq:banaszek} (see Appendix~\ref{asec:GDvsGDR}), which guarantees that $\cal G$-$\cal D$-$\cal R$ tightens $\cal G$-$\cal D$.

We here introduce and prove another useful inequality on the operation fidelity by a reversing operation as below:\\ 

\noindent
{\bf Lemma 2}. The overall operation fidelity covering all output states without postselection by the quantum measurement and the reversal ${\bf R}\circ{\bf M}$ is upper bounded by the operation fidelity of the quantum measurement $\bf M$, i.e.,
\begin{equation}
\label{eq:l2}
{\cal F}({\bf R}\circ{\bf M})\leq {\cal F}({\bf M}).
\end{equation}
%where ${\cal F}({\bf M})={\cal F}$ in our previous definition of Eq. (2).\\

See Appendix~\ref{asec:l2} for the details of the proof. Note that the equality in \eqref{eq:l2} holds when $\bf M$ is a unitary operation or a von Neumann measurement. The inequality \eqref{eq:l2} implies that `the disturbance in quantum measurement never decreases by any subsequent reversing operation.' %From the condition of the reversing operation, i.e., $({\bf R}\circ{\bf M})(\rho) \propto \rho$, the inequality is also valid for any subsequent measurement $\bf M'$ applied after $\bf M$, s.t.~${\cal F}({\bf M'}\circ{\bf M})\leq{\cal F}({\bf R}\circ{\bf M})\leq {\cal F}({\bf M})$. 
We can further generalize this for arbitrary $k$-times sequential quantum measurements, s.t.~${\cal F}({\bf M}_k\circ\cdots{\bf M}_2\circ{\bf M}_1)\leq\cdots\leq{\cal F}({\bf M}_2\circ{\bf M}_1)\leq {\cal F}({\bf M}_1)$. 

The inequality in \eqref{eq:l2} and its extension mentioned above are intuitively plausible by the second law of thermodynamics. These indicates the non-increasing of the average fidelity between the input and output states by reversing operations so that it differs from, but may be fundamentally related to, the data processing inequality \cite{Junge18}. We use \eqref{eq:l2} to derive a trade-off relation in what follows.

%%%%%%%%%%%%%%%%%%%%%%%%%%%%%%%%%%%%%%%
%%%%%%%%%%%%%%%%%%%%%%%%%%%%%%%%%%%%%%%%%%

Let us now derive the quantitative relation between the disturbance and the reversibility of quantum measurement using Lemma 1 and Lemma 2.\\

\noindent
{\bf Theorem 2}. ($\cal D$-$\cal R$: Disturbance and Reversibility trade-off) The disturbance ${\cal D}$ and the reversibility $\cal{R}$ of quantum measurements always satisfy
\begin{equation}
\label{eq:t2}
(d-1){\cal R}+(d+1){\cal D}\leq d-1,
\end{equation}
in an arbitrary dimension $d$.\\

{\em Proof}-- From Lemma 1 and Lemma 2, $(d-1){\cal R} \leq (d+1){\cal F}({\bf R}\circ{\bf M})-2 \leq (d+1){\cal F} -2$. By ${\cal D}=1-{\cal F}$, we obtain the inequality \eqref{eq:t2}. $\square$

The inequality in \eqref{eq:t2} complements other trade-off relations, determining the upper bound of $\cal{R}$ by ${\cal D}$. It implies that `the more disturbing a quantum measurement is, the less reversible it becomes.' Here the equality holds for von Neumann measurements or unitary operations. 

%%%%%%%%%%%%%%%%%%%%%%%%%%%%%%%%%%%%%%%
\begin{figure}[b]
\centering
\includegraphics[width=1.0\linewidth]{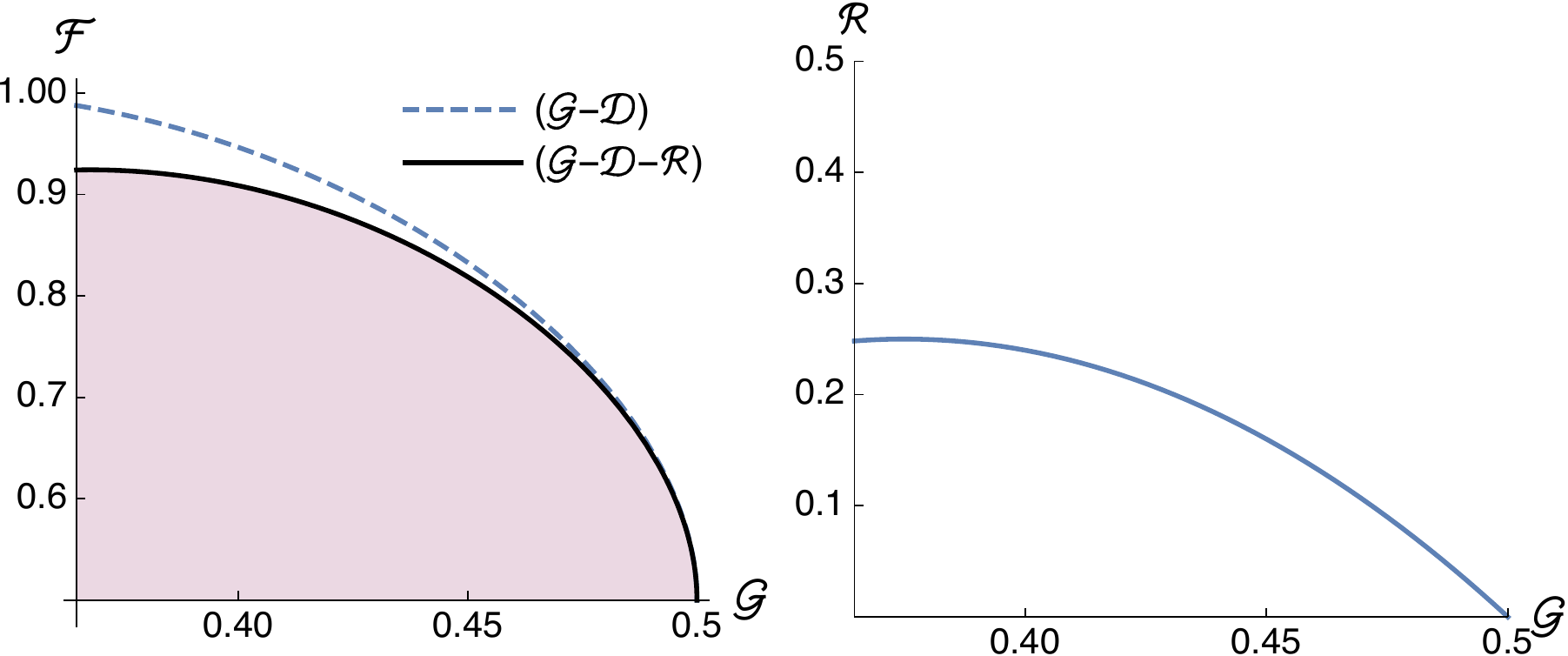}
\caption{Left: Comparison of the upper bounds by $\cal G$-$\cal D$-$\cal R$ and $\cal G$-$\cal D$ relations for the measurement in the main text. The bound by $\cal G$-$\cal D$-$\cal R$ is tighter than the one by $\cal G$-$\cal D$. Right: The gap is due to the reversibility $\cal R$, which is also in trade-off relation with the information gain $\cal G$.
}\label{fig:fig3}
\end{figure}
%%%%%%%%%%%%%%%%%%%%%%%%%%%%%%%%%%%%%%%

We have now completed the full quantitative links among the three information contents, i.e., information gain $\cal G$, disturbance $\cal D$, and reversibility $\cal R$ of quantum measurement, as illustrated in Fig.~\ref{fig:fig2}. 
These clearly show how the total information is balanced during the measurement and the reversal process. Suppose that we choose and modify a quantum measurement, intending to extract information maximally from the input quantum state. Such an optimization inevitably increases the disturbance $\cal D$ (i.e., decreases $\cal F$) from the relation $\cal G$-$\cal D$. The operation fidelity $\cal F$ can be interpreted here as the remaining information in the post-measurement state as a part of the initially encoded information into the input state. On the other hand, there exists a hidden part of information, by which we can recover the input state using a subsequent reversing operation on the output state, quantified by $\cal R$. 
It turns out that $\cal R$ fills the gap between $\cal G$ and $\cal D$ and tightens further the trade-off relations beyond $\cal G$-$\cal D$ as proved in this section. %Therefore, the total information contained initially in the input state is transferred to the {\em obtained information}, the {\em remaining information} at the post-measurement state, and the {\em reversible information}, which are associated with $\cal G$, $\cal F$, and $\cal R$, respectively. 

As an example, let us consider a quantum measurement with operators $\hat{M}_i=\sqrt{p}\ket{i}\bra{i}+\sqrt{(1-p)(3-p)/3}\ket{i+1}\bra{i+1}+\sqrt{p(1-p)/3}\ket{i+2}\bra{i+2}$ ($i=0,1,2$) for $0.458619\leq p\leq1$. The basis states are $\ket{i}\equiv\ket{i\mod3}$ in the Hilbert space of $d=3$. Using Eqs.~\eqref{eq:estifide},\eqref{eq:maxopfid} and \eqref{eq:revmax}, we can obtain the information gain ${\cal G}=(1+p)/4$, the operation fidelity ${\cal F}=1/4+(\sqrt{p}+\sqrt{(1-p)(3-p)/3}+\sqrt{p(1-p)/3})^2/4$, and the reversibility ${\cal R}=p(1-p)$. These turn out to saturate the global $\cal G$-$\cal D$-$\cal R$ tradeoff in Eq.~\eqref{eq:t1} but not the $\cal G$-$\cal D$ one in Eq.~\eqref{eq:banaszek}. In Fig.~\ref{fig:fig3} (left), we show the curve (black solid) representing the values of the pair ${\cal G}$ and ${\cal F}$ corresponding to the considered measurement, together with the curve (blue dashed) representing the bound from the $\cal G$-$\cal D$ relation in Eq.~\eqref{eq:banaszek}. We clearly see that the bound by $\cal G$-$\cal D$-$\cal R$ is tighter than the one by $\cal G$-$\cal D$. 

In view of the $\cal G$-$\cal D$ tradeoff, the measurement is not optimal in the sense that it does not maximize the output fidelity ${\cal F}$ for a given degree of ${\cal G}$. However, in fact, it is an optimal measurement saturating the $\cal G$-$\cal D$-$\cal R$ tradeoff in Eq.~\eqref{eq:t1}. The gap between the solid and the dashed curves is due to the reversibility $\cal R$, which is also in trade-off with $\cal G$ as plotted in Fig.~\ref{fig:fig3} (right). It thus illustrates the importance of including all information contents $\cal G$-$\cal D$-$\cal R$ to characterize a quantum measurement as optimal or non-optimal.

We note that there would exist some missing parts of the information accounted for by none of $\cal G$, $\cal F$, or $\cal R$. Such a missing part should be due to the non-optimality of quantum measurement or ignorance in the estimation of the input state based on the measurement outcomes. For instance, if we take into account the effect of errors in quantum measurement leading to imperfect reversal, the final output state may not be the same as the input state, i.e., $({\bf R}\circ{\bf M})(\rho) \propto \rho' \neq\rho$. Specifically, let us consider the case when the reversing operation succeeds but yields the output state $\rho(\psi,\epsilon)$ as
\begin{equation}
\hat{R}_{r,1}\hat{M}_r\ket{\psi}\bra{\psi}\hat{M}_r^\dag\hat{R}_{r,1}^\dag=|\eta_{r,1}|^2\rho(\psi,\epsilon)
\end{equation}
which deviates from the original input state due to an error parameterized by $0\leq \epsilon \leq 1$. The reversibility can then be evaluated as (see Appendix~\ref{asec:EE})
\begin{equation}
{\cal R}=\sum_r(\lambda^r_{d-1})^2 F_{s}.
\end{equation}
Here $F_{s}=\int d\psi \int^1_0 d\epsilon~p(\epsilon)\bra{\psi}\rho(\psi,\epsilon)\ket{\psi}$ is the average fidelity between the input state and the output state of the successful reversal, where $p(\epsilon)$ denotes the error probability density. %We assume here that the error occurs independently of the measurement outcome for simplicity. 
It shows that errors tend to decrease the reversibility in proportion to $F_{s}$. Note that when $F_s=1$ without errors the reversibility can reach the maximum value in Eq.~\eqref{eq:revmax}. 
Similarly, the amount of information gain can be degraded from the maximum $\cal G$ in Eq.~\eqref{eq:estifide} in the presence of errors. Therefore, imperfection and errors result in a missing part of the total information. As a result, the information contents $\cal G$, $\cal D$ and $\cal R$ of the quantum measurement with errors satisfy but cannot saturate the information trade-off relations.

In this context, we may define the {\em optimal quantum measurement} as the one that reaches the upper bounds of trade-off relations without any missing part of the total information. This further generalizes the definition of optimal quantum measurement to reach the bound of $\cal G$-$\cal D$ with minimal disturbance in \cite{Sacchi06,Sciarrino06,HTLim14}. In a sense, the optimal measurement is the one that conserves the total information in a nontrivial form obeying the trade-off relations. We present how optimal quantum measurements can be classified into different sets according to the considered trade-off relations in the next section.\\

%%%%%
%%%%%%%%%%%%%%%%%%%%%%%%%%%%%%%%%%%%%%%
\begin{figure}
\centering
\includegraphics[width=0.85\linewidth]{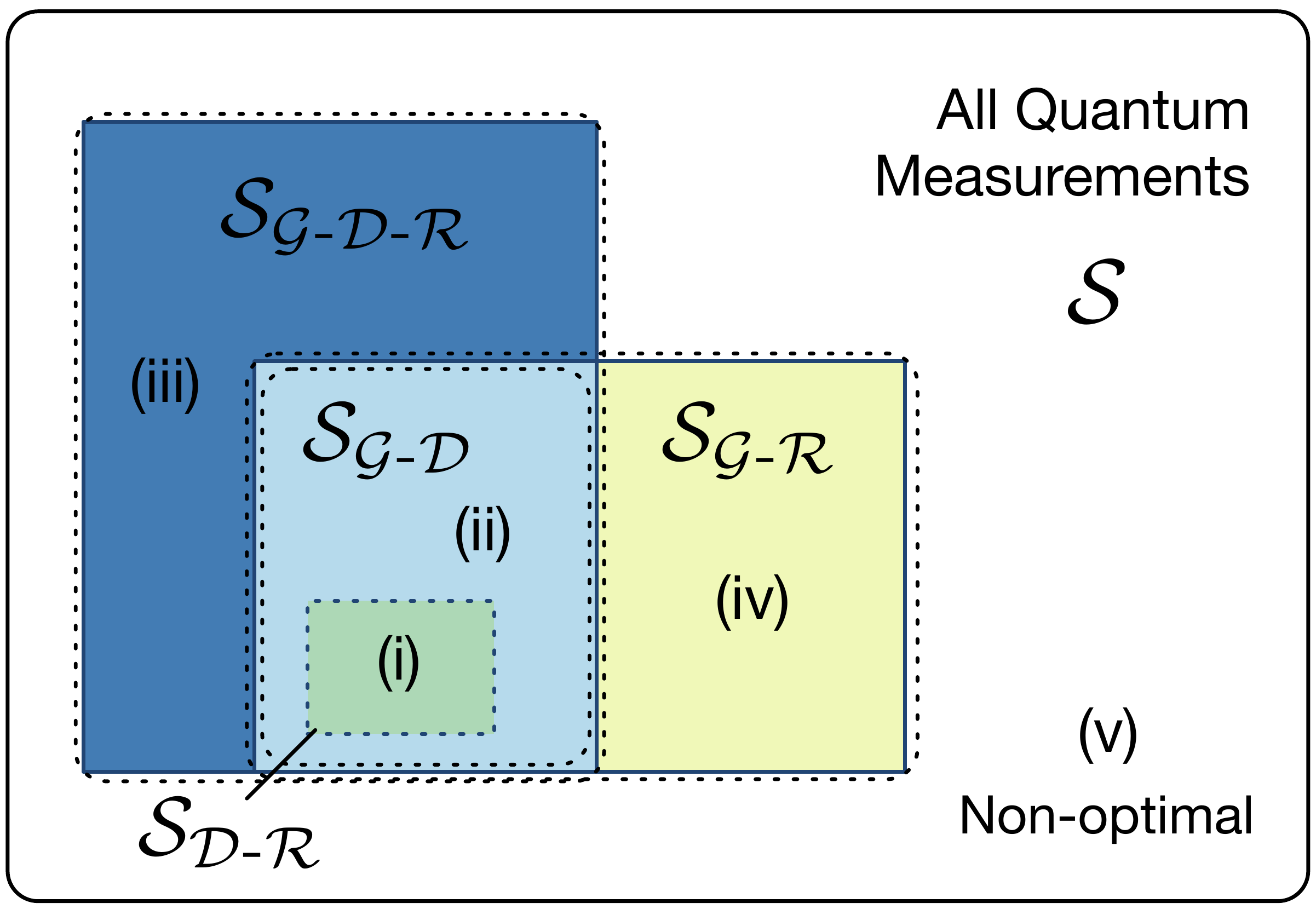}
\caption{A Venn diagram for quantum measurements based on the optimality to reach the bounds of $\cal G$, $\cal D$, and $\cal R$ by the trade-off relations. The quantum measurements (i)-(iv) presented in the main text are the representative examples of the different types of optimal quantum measurements in each divided region, while the measurement in the region (v) is non-optimal.
}\label{fig:fig4}
\end{figure}
%%%%%%%%%%%%%%%%%%%%%%%%%%%%%%%%%%%%%%%%%

%%%%%%%%%%%%%%%%%%%%%%%%%%%%%%%%%%%%%%
\begin{table}[b]
\caption{\label{tab:table1} Saturation condition on the trade-off relations.}
\begin{tabular}{l|c}
\hline\hline
&Saturation condition\\
\hline
$\cal G$-$\cal D$-$\cal R$ & all $\vec{v}_i$ are collinear\footnote{$\vec{v}_i=(\lambda^{r=1}_i, \ldots , \lambda^{r=n}_i)$} \& $|\vec{v}_1|= \cdots =|\vec{v}_{d-2}|$ \\
$\cal G$-$\cal D$ & all $\vec{v}_i$ are collinear \& $|\vec{v}_1|= \cdots =|\vec{v}_{d-1}|$ \\
$\cal G$-$\cal R$ & $\hat{M}^{\dag}_r\hat{M}_r=a_r\ket{j_r}\bra{j_r}+b_r\hat{\openone}$\footnote{$j_r \in \{0,\ldots,d-1\}$} with $a_r, b_r \geq 0$ \\
$\cal D$-$\cal R$ & von Neumann meas. or unitary op. \\
\hline\hline
\end{tabular}
\end{table}
%%%%%%%%%%%%%%%%%%%%%%%%%%%%%%%%%%%%%%%

\section{Classifying optimal quantum measurements}
\label{sec:Class}

%\subsection{Quantum measurement sets}

The saturation condition on each trade-off relation determines the criterion for classifying optimal quantum measurements. Let $\cal S$ be the set of all possible quantum measurements. We then denote by $\cal{S}_{\cal G\text{-}D\text{-}R}$, $\cal{S}_{\cal G\text{-}D}$, $\cal{S}_{\cal G\text{-}R}$, and $\cal{S}_{\cal D\text{-}R}$ those subsets of quantum measurements saturating the trade-off relations $\cal G$-$\cal D$-$\cal R$, $\cal G$-$\cal D$, $\cal G$-$\cal R$, and $\cal D$-$\cal R$, respectively. The saturation condition of $\cal G$-$\cal R$ is $\hat{M}^{\dag}_r\hat{M}_r=a_r\ket{j_r}\bra{j_r}+b_r\hat{\openone}$ where $j_r \in \{0,\cdots,d-1\}$ and $a_r$ and $b_r$ are non-negative parameters \cite{Cheong12}. It was shown in Ref.~\cite{HTLim14} that a quantum measurement saturating $\cal G$-$\cal D$ always satisfies the condition saturating $\cal G$-$\cal R$, while the converse is not true, s.t. $\cal{S}_{\cal G\text{-}R} \supset \cal{S}_{\cal G\text{-}D}$. In the previous section from Theorem 1, we have observed that $\cal{S}_{\cal G\text{-}D\text{-}R}\supset \cal{S}_{\cal G\text{-}D}$, i.e, a quantum measurement saturating $\cal G$-$\cal D$ always saturates $\cal G$-$\cal D$-$\cal R$, but the converse is not true. We further find that $\cal{S}_{\cal G\text{-}D\text{-}R} \cap \cal{S}_{\cal G\text{-}R}=\cal{S}_{\cal G\text{-}D}$ from the saturation conditions of $\cal G$-$\cal D$-$\cal R$ and $\cal G$-$\cal R$ (Table~\ref{tab:table1}). The detailed proof is given in Appendix~\ref{asec:SetP}. The elements of $\cal{S}_{\cal D\text{-}R}$ are either unitary operators or von Neumann measurements by Theorem 2. Now, the Venn diagram of optimal quantum measurement sets can be constructed as illustrated in Fig.~\ref{fig:fig4}, classified based on the optimality of quantum measurement to reach the upper bounds of information contents by the trade-off relations.

%\subsection{Examples}

Let us consider some examples of different types of quantum measurements.

(i) Assume that a von Neumann measurement $\hat{P}=\ket{i}\bra{i}$ is performed on arbitrary $d$-dimensional quantum states. It allows one to obtain the maximum information ${\cal G}=2/(d+1)$. Therefore, no significant information remains on the post-measurement state ${\cal F}=2/(d+1)$ nor is recoverable ${\cal R}=0$ (we remind that the range of information gain, operation fidelity, and reversibility are $1/d \leq {\cal G} \leq 2/(d+1)$, $2/(d+1)\leq {\cal F} \leq 1$, and $0\leq {\cal R} \leq 1$, respectively). It is straightforward to see that these quantities saturate all the trade-off relations (Fig.~\ref{fig:fig4}). 

(ii) Consider a quantum measurement $\hat{M}_i=\sqrt{p}\ket{i}\bra{i}+\sqrt{(1-p)/2}(\hat{\openone}-\ket{i}\bra{i})$ ($i=0,1,2$) for $1/3\leq p\leq1$. It becomes a von Neumann measurement when $p=1$ and a unitary operator when $p=1/3$. It causes a partial collapse of the input state and can be reversed for $1/3< p<1$. For each outcome $i$, the optimal reversing operation is defined by $\hat{R}_{i,1}=\sqrt{(1-p)/2p}\ket{i}\bra{i}+(\hat{\openone}-\ket{i}\bra{i})$ (success) and $\hat{R}_{i,2}=\sqrt{(3p-1)/2p}\ket{i}\bra{i}$ (non-success). We obtain the information contents as ${\cal G}=(1+p)/4$, ${\cal F}=(3-p+2\sqrt{2p(1-p)})/4$, and ${\cal R}=3(1-p)/2$, which are saturating $\cal G$-$\cal D$, $\cal G$-$\cal D$-$\cal R$, and $\cal G$-$\cal R$ relations, i.e.,~$\{\hat{M}_r\}\in S_{\cal G\text{-}D\text{-}R}$, $\{\hat{M}_r\}\in S_{\cal G\text{-}D}$ and $\{\hat{M}_r\}\in S_{\cal G\text{-}R}$.

(iii) Consider a quantum measurement $\hat{M}_i=\sqrt{p}\ket{i}\bra{i}+\sqrt{2(1-p)/3}\ket{i+1}\bra{i+1}+\sqrt{(1-p)/3}\ket{i+2}\bra{i+2}$ ($i=0,1,2$) for $2/5\leq p\leq1$ with $\ket{i}\equiv\ket{i\mod3}$ in 3-dimensional Hilbert space.
It can be optimally reversed by $\hat{R}_{i,1}=\sqrt{(1-p)/3p}\ket{i}\bra{i}+\sqrt{1/2}\ket{i+1}\bra{i+1}+\ket{i+2}\bra{i+2}$ (success) and $\hat{R}_{i,2}=\sqrt{(4p-1)/3p}\ket{i}\bra{i}+\sqrt{1/2}\ket{i+1}\bra{i+1}$ (non-success). We obtain ${\cal G}=(1+p)/4$, ${\cal F}=(3+\sqrt{2}(1-p)+(\sqrt{3}+\sqrt{6})\sqrt{p(1-p)})/6$, and ${\cal R}=1-p$. These saturate $\cal G$-$\cal D$-$\cal R$ but do not saturate $\cal G$-$\cal D$ and $\cal G$-$\cal R$, i.e.,~$\{\hat{M}_r\}\in S_{\cal G\text{-}D\text{-}R}$, $\{\hat{M}_r\}\notin S_{\cal G\text{-}D}$ and $\{\hat{M}_r\}\notin S_{\cal G\text{-}R}$.

(iv) Consider a quantum measurement $\hat{M}_i=\sqrt{1/3}\ket{i}\bra{i}+\sqrt{p/6}(\ket{i+1}\bra{i+1}+\ket{i+2}\bra{i+2})$ ($i=0,1$) and $\hat{M}_2=\sqrt{(3-p)/3}\ket{2}\bra{2}+\sqrt{(4-p)/6}(\ket{0}\bra{0}+\ket{1}\bra{1})$ for $0\leq p\leq1$. It can be optimally reversed by $\hat{R}_{i,1}=\sqrt{p/2}\ket{i}\bra{i}+\ket{i+1}\bra{i+1}+\ket{i+2}\bra{i+2}$ (success) and $\hat{R}_{i,2}=\sqrt{(1-p)/2}\ket{i}\bra{i}$ (non-success) for $i=0,1$, and $\hat{R}_{2,1}=\sqrt{(4-p)/2(3-p)}\ket{2}\bra{2}+\ket{0}\bra{0}+\ket{1}\bra{1}$ (success) and $\hat{R}_{2,2}=\sqrt{(2-p)/2(3-p)}\ket{2}\bra{2}$ (non-success).
We obtain ${\cal G}=(14-p)/36$, ${\cal F}=(32-p+4\sqrt{2p}+2\sqrt{6-2p}\sqrt{4-p})/72$, and ${\cal R}=(4+p)/6$. Notably, these saturate $\cal G$-$\cal R$ but do not saturate $\cal G$-$\cal D$-$\cal R$ and $\cal G$-$\cal D$, i.e.,~$\{\hat{M}_r\}\in S_{\cal G\text{-}R}$, $\{\hat{M}_r\}\notin S_{\cal G\text{-}D\text{-}R}$ and $\{\hat{M}_r\}\notin S_{\cal G\text{-}D}$.

(v) Consider a weak quantum measurement with operators $\hat{M}_1=\sqrt{p}\ket{1}\bra{1}$ and $\hat{M}_2=\ket{0}\bra{0}+\sqrt{1-p}\ket{1}\bra{1}+\ket{2}\bra{2}$, performed on an arbitrary quantum state $\ket{\psi}$ in 3-dimensional Hilbert space. While the input state $\ket{\psi}$ is completely collapsed on $\ket{1}$ when the outcome $r=1$, it is partially collapsed when $r=2$ so that the measurement is reversible for $p<1$. The optimal reversing operators are given by $\hat{R}_{2,1}=\sqrt{1-p}\ket{0}\bra{0}+\ket{1}\bra{1}+\sqrt{1-p}\ket{2}\bra{2}$ (success) and $\hat{R}_{2,2}=\sqrt{p}\ket{0}\bra{0}+\sqrt{p}\ket{2}\bra{2}$ (non-success). The amount of obtained information is ${\cal G}=(4+p)/12$, while the remaining and the reversible ones are ${\cal F}=(2+\sqrt{1-p})/3$ and ${\cal R}=1-p$, respectively. While these quantities satisfy all the trade-off relations, they do not saturate any of them.

We have observed that all of the above examples satisfy the trade-off inequalities. The examples (i) - (iv) are optimal quantum measurements that saturate at least one of the trade-off relations. In an optimal quantum measurement, the total information is divided into $\cal G$, $\cal D$, and $\cal R$ and balanced by the change of the parameter $p$. On the other hand, the example (v) is non-optimal which satisfies all the trade-off inequalities but does not saturate any of them. Each of the examples (i) - (v) given above represents each divided region in the diagram of Fig.~\ref{fig:fig4}. 

\section{Discussion}

We have established the complete information balance in quantum measurement by deriving the full quantitative trade-off relations among information gain, disturbance, and reversibility. 
Under a quantum measurement, the initial information contained in the input state is divided into i) the obtained information $\cal G$, ii) the remaining information $\cal F$ in the post-measurement state, and iii) the reversible information $\cal R$. 
Our result clearly shows that the three quantities are balanced by the trade-off relations. The reversibility $\cal R$ turns out to play an essential role in completing the information balance, filling the gap between the information gain $\cal G$ and disturbance ${\cal D}=1-{\cal F}$. Note that the three information contents are defined to be universal for a given dimension by covering all possible input states and have clear operational meanings, fulfilling a general requirement as an information content in quantum measurement \cite{Berta14}. 

While all quantum measurements, including noisy or weak measurements, must satisfy the trade-off relations, the conditions to saturate them define optimal quantum measurements. Those optimal measurements may be said to conserve the total information in a nontrivial form according to the trade-off relation. We have classified all quantum measurements into different sets based on their optimality to reach the upper bounds of $\cal G$, $\cal D$, and $\cal R$. We then thoroughly analyzed their relations resulting in the Venn diagram in Fig.~\ref{fig:fig4}. These may provide useful guidelines for designing a quantum measurement according to its aim in quantum information protocol. For example, a maximum information gain with minimal disturbance is desirable for estimating or discriminating quantum information \cite{HTLim14}, whereas a maximum reversibility suits the aim of transmitting \cite{AdvancedTele} or protecting \cite{QEC} quantum information. %Details of the applications and implementations of the optimal quantum measurements will be presented elsewhere \cite{Hong2020}.

Our results imply that total information does not increase in quantum measurement and reversal process and may hint at the extension of the information conservation law. It should be obviously the case with a unitarity process in line with the second law of thermodynamics \cite{Buscemi16,Kwon19}. Extending to the non-unitary quantum measurement and selective processes, our optimal measurements saturating trade-off relations may be considered as defining the information conservation. 

\begin{figure}
\centering
\includegraphics[width=0.8\linewidth]{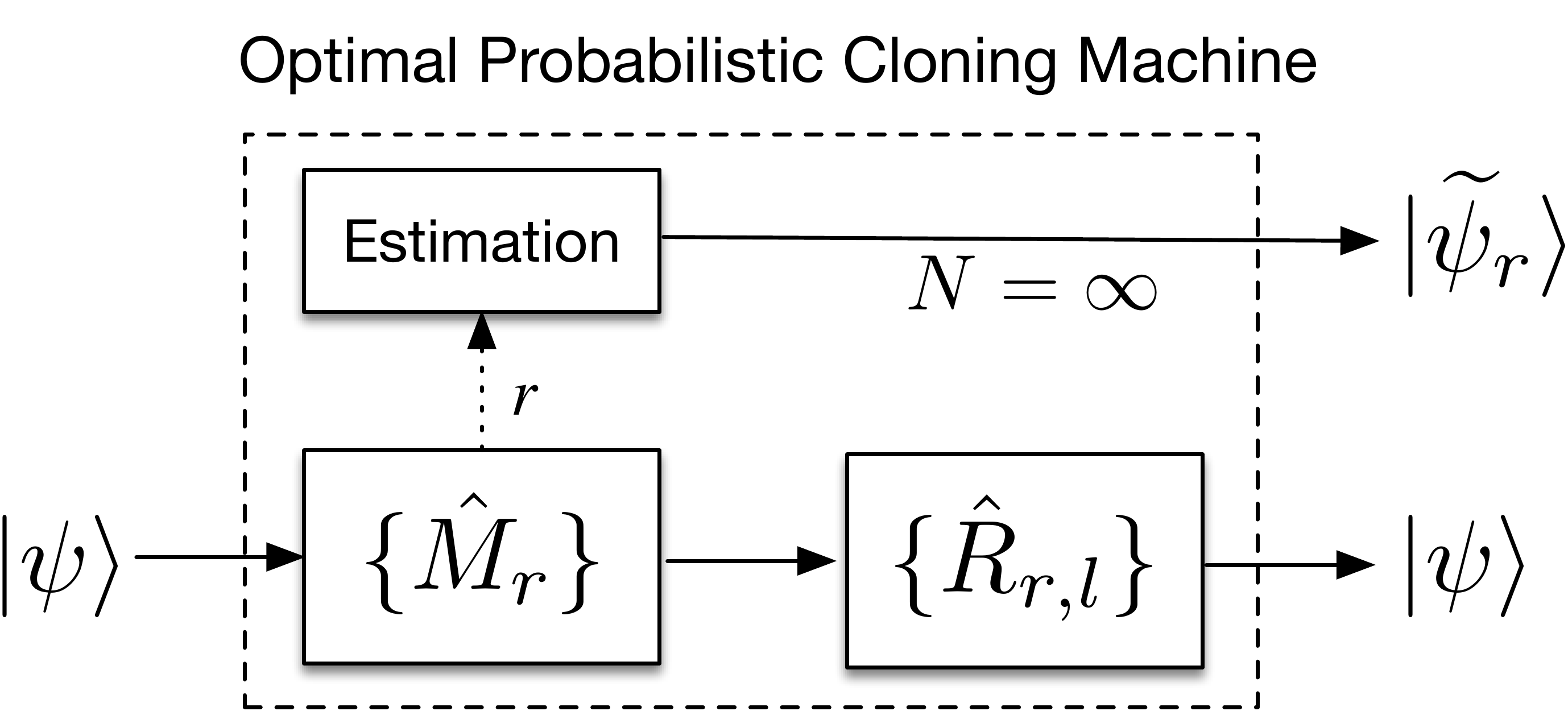}
\caption{The state estimation by a quantum measurement $\{\hat{M}_r\}$ followed by a reversal process is equivalent to an $1\rightarrow N+1$ asymmetric probabilistic cloning machine as $N\rightarrow\infty$. The output at the bottom yields exactly the same state as the input state with the probability given by $\cal R$, while the upper estimation part yields an infinite number of approximated copies with the fidelity given by $\cal G$.
}\label{fig:fig5}
\end{figure}

Our results can also be interpreted as a quantitative refinement of the no-cloning theorem \cite{NoClon} in the context of quantum measurement. As a direct application, we can consider a universal cloning machine, which has been analyzed so far either as a deterministic process with a lower fidelity \cite{ClonM1,ClonM2,ClonM3,Bruss98,Iblisdir05,Bae06} or a probabilistic one for an exact cloning \cite{Duan98l,Duan98a}. By contrast, our results allow us to optimize further the protocol by compromising the output fidelity and the success probability of the cloning. For example, consider a $1\rightarrow N+1$ asymmetric, probabilistic, cloning machine yielding
\begin{equation}
\ket{\psi}\rightarrow\eta_{r,l}\ket{\tilde{\psi}_r}^{\otimes N}\ket{\psi},
\end{equation}
where $\ket{\tilde{\psi}_r}^{\otimes N}$ are the approximate $N$ copies and $\ket{\psi}$ a perfect copy.
Employing the optimal measurement and the reversal process can realize this cloning machine when $N\rightarrow\infty$, as illustrated in Fig.~\ref{fig:fig5}, with $|\eta_{r,l}|^2$ the success probability, and $r$ and $l$ the outcomes of the measurement and the reversal process, respectively.  Compared to the deterministic version \cite{Bruss98,Iblisdir05,Bae06}, our results provide a method to optimize the cloning process further with enhanced fidelities at the expense of success probability. Note that the trade-off relations among $\cal G$, $\cal D$, and $\cal R$ determine the quantitative upper bound of the performance of the cloning machine.

%%%%%%%%%%%%%%%%%%%%%%%%%%%%%%%%%%%%%%%
%%%%%%%%%%%%%%%%%%%%%%%%%%%%%%%%%%%%%%%%%%

An important path for further study is exploring practical applications, e.g. measurement-based quantum processor, teleportation, metrology, and quantum error correction. Characterizing the information flow in sequential quantum measurements would also be interesting, in which the uncertainty relation between incompatible measurements may be crucial. It may also be valuable to translate our results into the context of quantum thermodynamics \cite{Sagawa08,Jacobs09}. As the information contents $\cal G$, $\cal D$, and $\cal R$ are defined as measurable quantities, the derived trade-off relations are ready to be tested in any quantum information platform, e.g., superconducting \cite{Katz08}, ion trap \cite{Schindler13}, and photonic qubits \cite{YSKim09,YSKim11,JCLee11,HTLim14,Chen14}. Our work establishes fundamental criteria for characterizing quantum measurement and offers useful guidelines for designing optimal measurement-based quantum processors. 

\acknowledgments
This work was supported in part by the KIAS Advanced Research Program (QP029902 and CG014604), the National Research Foundation of Korea (NRF) (2020M3E4A1079939), and the KIST institutional program (2E31021).

\bibliographystyle{plainnat}

\begin{thebibliography}{9}
  
\bibitem{Heisenberg} W. Heisenberg, {\it {\"U}ber den anschaulichen inhalt der quantentheoretischen kinematik und mechanik}, 
\href{https://doi.org/10.1007/BF01397280}
{Z. Phys. {\bf 43}, 172--198 (1927).}
\bibitem{QN1} M. T. Quintino, T. V\'{e}rtesi, and N. Brunner, Joint Measurability, Einstein-Podolsky-Rosen Steering, and Bell Nonlocality, \href{https://doi.org/10.1103/PhysRevLett.113.160402}{Phys. Rev. Lett. {\bf 113}, 160402 (2014).}
\bibitem{QN2} R. Uola, T. Moroder, and O. G\"{u}hne, Joint Measurability of Generalized Measurements Implies Classicality, 
\href{https://doi.org/10.1103/PhysRevLett.113.160403}
{Phys. Rev. Lett. {\bf 113}, 160403 (2014).}
\bibitem{OQ} R. Raussendorf and H. J. Briegel, A One-Way Quantum Computer,
\href{https://doi.org/10.1103/PhysRevLett.86.5188}
{Phys. Rev. Lett. {\bf 86}, 5188 (2001).}
\bibitem{QIQM} M. Nielsen and I. Chuang, 
\href{https://doi.org/10.1017/CBO9780511976667}
{{\em Quantum Computation and Quantum Information} (Cambridge University Press, 2000).}
\bibitem{Wiseman09} H.~M. Wiseman and G.~J. Milburn, 
\href{https://doi.org/10.1017/CBO9780511813948}
{{\em Quantum Measurement and Control} (Cambridge University Press, 2009).}
\bibitem{Kurt} K. Jacobs,
\href{https://doi.org/10.1017/CBO9781139179027}
{{\em Quantum Measurement Theory and its Applications} (Cambridge University Press, 2014).}
\bibitem{teleportation} C. H. Bennett, G. Brassard, C. Cr\'epeau, R. Jozsa, A. Peres, and W. K. Wootters, Teleporting an unknown quantum state via dual classical and einstein-podolsky-rosen channels, 
\href{https://doi.org/10.1103/PhysRevLett.70.1895}
{Phys. Rev. Lett. {\bf 70}, 1895 (1993).}
\bibitem{MBQC} H. J. Briegel, D. E. Browne, W. D\"ur, R. Raussendorf, and M. Van den Nes, Measurement-based quantum computation, 
\href{https://doi.org/10.1038/nphys1157}
{Nature Physics {\bf 5}, 19--26 (2009).}
\bibitem{AdvancedTele} S. Pirandola, J. Eisert, C. Weedbrook, A. Furusawa, and S. L. Braunstein, Advances in quantum teleportation, 
\href{https://doi.org/10.1038/nphoton.2015.154}
{Nature Photonics {\bf 9}, 641--652 (2015).}
\bibitem{AdvancedMetro} V. Giovannetti, S. Lloyd, and L. Maccone, Advances in quantum metrology, 
\href{https://doi.org/10.1038/nphoton.2011.35}
{Nature Photonics {\bf 5}, 222--229 (2011).}
\bibitem{QEC} D. Lidar and T. Brun, 
\href{https://doi.org/10.1017/CBO9781139034807}
{{\em Quantum Error Correction} (Cambridge University Press, 2013).}
 \bibitem{Gisin2002} N. Gisin, G. Ribordy, W. Tittel, and H. Zbinden, Quantum cryptography, 
\href{https://doi.org/10.1103/RevModPhys.74.145}
{Rev. Mod. Phys. {\bf 74}, 145 (2002).}
 \bibitem{BB84} C. H. Bennett and G. Brassard, {\it Quantum cryptography: Public key distribution and coin tossing}, 
 \href{https://doi.org/10.1016/j.tcs.2014.05.025}
{Proceedings of IEEE International Conference on Computers, Systems and Signal Processing, 175 (1984).}
\bibitem{Groenewold71} H. J. Groenewold, A problem of information gain by quantal measurements, 
\href{https://doi.org/10.1007/BF00815357}
{Int. J. Theoretical Phys. {\bf 4}, 327--338 (1971).}
\bibitem{Lindblad72}  G. Lindblad, An entropy inequality for quantum measurements, 
\href{https://doi.org/10.1007/BF01645778}
{Commun. Math. Phys. {\bf 28}, 245--249 (1972).}
\bibitem{Ozawa86} M. Ozawa, On information gain by quantum measurements of continuous observables, 
\href{https://doi.org/10.1063/1.527179}
{J. Math. Phys. {\bf 27}, 759 (1986).}
\bibitem{Fuchs96} C. A. Fuchs and A. Peres, Quantum-state disturbance versus information gain: Uncertainty relations for quantum information, 
\href{https://doi.org/10.1103/PhysRevA.53.2038}
{Phys. Rev. A {\bf 53}, 2038 (1996).}
\bibitem{Fuchs01} C. A. Fuchs and K. A. Jacobs, Information-tradeoff relations for finite-strength quantum measurements, 
\href{https://doi.org/10.1103/PhysRevA.63.062305}
{Phys. Rev. A {\bf 63}, 062305 (2001).}
\bibitem{Banaszek01} K. Banaszek, Fidelity balance in quantum operations,
\href{https://doi.org/10.1103/PhysRevLett.86.1366}
{Phys. Rev. Lett. {\bf 86}, 1366 (2001).}
\bibitem{Banaszek02} K. Banaszek and I. Devetak, Fidelity trade-off for finite ensembles of identically prepared qubits, 
\href{https://doi.org/10.1103/PhysRevA.64.052307}
{Phys. Rev. A {\bf 64}, 052307 (2001).}
 \bibitem{Dariano03} G. M. D'ariano, On the Heisenberg principle, namely on the  information-disturbance trade-off in a quantum measurement, 
 \href{https://doi.org/10.1002/prop.200310045}
{Fortschr. Phys. {\bf 51}, 318 (2003).}
\bibitem{Sacchi06} M. F. Sacchi, Information-disturbance tradeoff in estimating a
  maximally entangled state, 
 \href{https://doi.org/10.1103/PhysRevLett.96.220502}
 {Phys. Rev. Lett. {\bf 96}, 220502 (2006).}
\bibitem{Buscemi08} F. Buscemi, M. Hayashi, and M. Horodecki, Global information balance in quantum measurements, 
\href{https://doi.org/10.1103/PhysRevLett.100.210504}
{Phys. Rev. Lett. {\bf 100}, 210504 (2008)}.
\bibitem{Luo10} S. Luo, Information conservation and entropy change in
  quantum measurements, 
\href{https://doi.org/10.1103/PhysRevA.82.052103}
{Phys. Rev. A {\bf 82}, 052103 (2010).}
\bibitem{Berta14} M. Berta, J. M. Renes, and M. M. Wilde, Identifying the information gain of a quantum measurement, 
\href{https://doi.org/10.1109/TIT.2014.2365207}
{IEEE Transactions on Information Theory {\bf 60}, 7987 (2014).}
\bibitem{Ueda92} M. Ueda and M. Kitagawa, Reversibility in quantum measurement processes, 
\href{https://doi.org/10.1103/PhysRevLett.68.3424}
{Phys. Rev. Lett. {\bf 68}, 3424 (1992).}
\bibitem{Royer95} A. Royer, Reversible quantum measurements on a spin 1/2 and measuring the state of a single system, 
\href{https://doi.org/10.1103/PhysRevLett.73.913}
{Phys. Rev. Lett. {\bf 73}, 913 (1994).}
\bibitem{Ueda96} M. Ueda, N. Imoto, and H. Nagaoka, Logical reversibility in quantum measurement: General theory and specific examples, 
\href{https://doi.org/10.1103/PhysRevA.53.3808}
{Phys. Rev. A {\bf 53}, 3808 (1996).}
\bibitem{Jordan10} A. N. Jordan and A. N. Korotkov, Uncollapsing the wavefunction by undoing quantum measurements, 
\href{https://doi.org/10.1080/00107510903385292}
{Contemporary Physics {\bf 51}, 125 (2010).}
\bibitem{Koashi99} M. Koashi and M. Ueda, Reversing measurement and probabilistic quantum error correction, 
\href{https://doi.org/10.1103/PhysRevLett.82.2598}
{Phys. Rev. Lett. {\bf 82}, 2598 (1999).}
\bibitem{Terashima03} H. Terashima and M. Ueda, Nonunitary quantum circuit,
\href{https://doi.org/10.1142/S0219749905001456}
{International Journal of Quantum Information {\bf 3}, 633 (2005).}
\bibitem{Korotkov06} A. N. Korotkov and A. N. Jordan, Undoing a weak quantum measurement of a solid-state qubit, 
\href{https://doi.org/10.1103/PhysRevLett.97.166805}
{Phys. Rev. Lett. {\bf 97}, 166805 (2006).}
\bibitem{Korotkov10} A. N. Korotkov and K. Keane, Decoherence suppression by quantum measurement reversal, 
\href{https://doi.org/10.1103/PhysRevA.81.040103}
{Phys. Rev. A {\bf 81}, 040103(R) (2010).}
\bibitem{YSKim09} Y.-S. Kim, Y.-W. Cho, Y.-S. Ra, and Y.-H. Kim, Reversing the weak quantum measurement for a photonic qubit, 
\href{https://doi.org/10.1364/OE.17.011978}
{Optics Express {\bf 17}, 11978 (2009).}
\bibitem{YSKim11} Y.-S. Kim, J.-C. Lee, O. Kwon, and Y.-H. Kim, Protecting entanglement from decoherence using weak measurement and quantum measurement reversal, 
\href{https://doi.org/10.1038/nphys2178}
{Nature Physics {\bf 8}, 117 (2012).}
\bibitem{Katz08} N. Katz, M. Neeley, M. Ansmann, R. C. Bialczak, M. Hofheinz, E. Lucero, A. O'Connell, H. Wang, A. N. Cleland, J. M. Martinis, and A. N. Korotkov, Reversal of the weak measurement of a quantum state in a superconducting phase qubit, 
\href{https://doi.org/10.1103/PhysRevLett.101.200401}
{Phys. Rev. Lett. {\bf 101}, 200401 (2008).}
\bibitem{Schindler13} P. Schindler, T. Monz, D. Nigg, J. T. Barreiro, E. A. Martinez, M. F. Brandl, M. Chwalla, M. Hennrich, and R. Blatt, Undoing a Quantum Measurement, 
\href{https://doi.org/10.1103/PhysRevLett.110.070403}
{Phys. Rev. Lett. {\bf 110}, 070403 (2013).}
\bibitem{JCLee11} J.-C. Lee, Y.-C. Jeong, Y.-S. Kim, and Y.-H. Kim, Experimental demonstration of decoherence suppression via quantum measurement reversal, 
\href{https://doi.org/10.1364/OE.19.016309}
{Optics Express {\bf 19}, 16309 (2011).}
\bibitem{Cheong12} Y. W. Cheong and S.-W. Lee, Balance between information gain and reversibility in weak measurement, 
\href{https://doi.org/10.1103/PhysRevLett.109.150402}
{Phys. Rev. Lett. {\bf 109}, 150402 (2012).}
\bibitem{HTLim14} H.-T. Lim, Y.-S. Ra, K.-H. Hong, S.-W. Lee, and Y.-H. Kim, Fundamental bounds in measurements for estimating quantum states, 
\href{https://doi.org/10.1103/PhysRevLett.113.020504}
{Phys. Rev. Lett. {\bf 113}, 020504 (2014).}
\bibitem{Chen14} G. Chen, Y. Zou, X.-Y. Xu, J.-S. Tang, Y.-L. Li, J.-S. Xu, Y.-J. Han, C.-F. Li, G.-C. Guo, H.-Q. Ni, Y. Yu, M.-F. Li, G.-W. Zha, Z.-C. Niu, and Y. Kedem, Experimental Test of the State Estimation-Reversal Tradeoff Relation in General Quantum Measurements, 
\href{https://doi.org/10.1103/PhysRevX.4.021043}
{Phys. Rev. X {\bf4}, 021043 (2014).}
\bibitem{Terashima11} H. Terashima, Information, fidelity, and reversibility in
  single-qubit measurements, 
 \href{https://doi.org/10.1103/PhysRevA.83.032114}
 {Phys. Rev. A {\bf 83}, 032114 (2011).}
\bibitem{Terashima16} H. Terashima, Information, fidelity, and reversibility in
  general quantum measurements, 
  \href{https://doi.org/10.1103/PhysRevA.93.022104}
  {Phys. Rev. A {\bf 93}, 022104 (2016).}
  
\bibitem{Petz86} D. Petz, Sufficient subalgebras and the relative entropy of
  states of a von Neumann algebra, 
  \href{https://doi.org/10.1007/BF01212345}
  {Commun. Math. Phys. {\bf 105}, 123 (1986).}
\bibitem{Petz88} D. Petz, Sufficiency of channels over von Neumann algebras, 
\href{https://doi.org/10.1093/qmath/39.1.97}
{The Quarterly Journal of Mathematics {\bf 39}, 97(1988).}
  
\bibitem{Junge18} M. Junge, R. Renner, D. Sutter, M. M. Wilde, and A. Winter, Universal recovery maps and approximate sufficiency of quantum relative entropy,
\href{https://doi.org/10.1007/s00023-018-0716-0}
{Annales Henri Poincare {\bf 19}, 2955 (2018).}
\bibitem{Sciarrino06} F. Sciarrino, M. Ricci, F. De Martini, R. Filip, and L. Mi{\v s}ta, Jr., Realization of a minimal disturbance quantum measurement, 
\href{https://doi.org/10.1103/PhysRevLett.96.020408}
{Phys. Rev. Lett. {\bf 96}, 020408 (2006).}

\bibitem{Buscemi16} F. Buscemi, S. Das, and M. M. Wilde, Approximate reversibility in the context of entropy gain, information gain, and complete positivity, 
\href{https://doi.org/10.1103/PhysRevA.93.062314}
{Phys. Rev. A {\bf 93}, 062314 (2016).}
\bibitem{Kwon19} H. Kwon and M. S. Kim, Fluctuation theorems for a quantum channel, 
\href{https://doi.org/10.1103/PhysRevX.9.031029}
{Phys. Rev. X {\bf 9}, 031029 (2019).}
\bibitem{NoClon} W. Wootters and W. Zurek, A Single Quantum Cannot be Cloned,
\href{https://doi.org/10.1038/299802a0}
{Nature {\bf 299}, 802--803 (1982).}

%\bibitem{Hong2020} S. Hong, Y.-S. Kim, Y.-W. Cho, J. Kim, S.-W. Lee, and H.-T. Lim, {\it in preparation}.
\bibitem{ClonM1} V. Bu{\v z}ek and M. Hillery, Quantum copying: Beyond the no-cloning theorem, 
\href{https://doi.org/10.1103/PhysRevA.54.1844}
{Phys. Rev. A {\bf 54}, 1844 (1996).}
\bibitem{ClonM2} N. Gisin and S. Massar, Optimal quantum cloning machines, 
\href{https://doi.org/10.1103/PhysRevLett.79.2153}
{Phys. Rev. Lett. {\bf 79}, 2153 (1997).}
\bibitem{ClonM3} V. Scarani, S. Iblisdir, N. Gisin, and A. Ac{\' i}n, Quantum cloning, 
\href{https://doi.org/10.1103/RevModPhys.77.1225}
{Rev. Mod. Phys. {\bf 77}, 1225 (2005).}
\bibitem{Bruss98} D. Bruss, A. Ekert, and C. Macchiavello, Optimal universal quantum cloning and state estimation, 
\href{https://doi.org/10.1103/PhysRevLett.81.2598}
{Phys. Rev. Lett. {\bf 81}, 2598 (1998).}
\bibitem{Iblisdir05} S. Iblisdir, A. Ac{\' i}n, N. J. Cerf, R. Filip, J. Fiur{\' a}{\v s}ek, and N. Gisin, Multipartite asymmetric quantum cloning, 
\href{https://doi.org/10.1103/PhysRevA.72.042328}
{Phys. Rev. A {\bf 72}, 042328 (2005).}
\bibitem{Bae06} J. Bae and A. Ac{\' i}n, Asymptotic quantum cloning is state estimation, 
\href{https://doi.org/10.1103/PhysRevLett.97.030402}
{Phys. Rev. Lett. {\bf 97}, 030402(2006).}
\bibitem{Duan98l} L. M. Duan and G. C. Guo, Probabilistic cloning and identification of linearly independent quantum states, 
\href{https://doi.org/10.1103/PhysRevLett.80.4999}
{Phys. Rev. Lett. {\bf 80}, 4999 (1998).}
\bibitem{Duan98a} L. M. Duan and G. C. Guo, A probabilistic cloning machine for replicating two non-orthogonal states, 
\href{https://doi.org/10.1016/S0375-9601(98)00287-4}
{Phys. Lett. A. {\bf 243}, 261 (1998).}
\bibitem{Sagawa08} T. Sagawa and M. Ueda, Second law of thermodynamics with discrete quantum feedback control, 
\href{https://doi.org/10.1103/PhysRevLett.100.080403}
{Phys. Rev. Lett. {\bf 100}, 080403 (2008).}
\bibitem{Jacobs09} K. Jacobs, Second law of thermodynamics and quantum feedback control: Maxwell's demon with weak measurements, 
\href{https://doi.org/10.1103/PhysRevA.80.012322}
{Phys. Rev. A {\bf 80}, 012322 (2009).}

%\bibitem{Niu98} C. S. Niu and R. B. Griffiths, Phys. Rev. A {\bf 58}, 4377 (1998).
%\bibitem{Cerf2000} N. J. Cerf, Phys. Rev. Lett. {\bf 84}, 4497 (2000).
%\bibitem{Buzek98} V. Bu{\v z}ek, M. Hillery, and R. Bednik, Acta Phys. Slovaca {\bf 48}, 177 (1998).
%\bibitem{Braunstein01} S. L. Braunstein, V. Bu{\v z}ek, and M. Hillery, Phys. Rev. A {\bf 63}, 052313 (2001).
%\bibitem{Cerf2002} N. J. Cerf, M. Bourennane, A. Karlsson, and N. Gisin, Phys. Rev. Lett. {\bf 88}, 127902 (2002).
%\bibitem{MBQ} H. J. Briegel, D. E. Browne, W. D\"ur, R. Raussendorf, and M. Van den Nest, Nature Physics {\bf5} 19--26 (2009).

\end{thebibliography}

%%%%%%%%%%%%%%%%%%%%%%%%%%%%%%%%%%%%%%%%%%%%%%%%%%%%%%%%%%%%%%%%%%%%%%%%%%%%%%%%%%%%%%%%%%%%%%%%%%%%%%%%%%%%%%%%%%%%%%%%
\onecolumn\newpage
\appendix

\section{Definitions}
\label{asec:Def}

\subsection{Singular value decomposition}
\label{asec:SVD}

By a singular value decomposition, a measurement operator $\hat{M}_r$ can be represented as $\hat{M}_r=\hat{V}_r\hat{D}_r\hat{W}_r$ in terms of unitary operators $\hat{W}_r$ and $\hat{V}_r$ and a diagonal matrix $\hat{D}_r=\sum_{i=0}^{d-1}\lambda^r_i\ket{i} \bra{i}$. Without loss of generality, we set $\hat{W}_r=\hat{\openone}$ and also assume that the singular values are defined here in decreasing order, $\lambda^r_0\geq \lambda^r_1\geq\ldots \geq\lambda^r_{d-1}\geq0$. Note that the singular values satisfy the completeness relation 
\begin{equation}
\label{eq:comp1st}
\sum_r\sum_i (\lambda^r_i)^2 =d.
\end{equation}
Similarly, the operator of reversing operation can be represented as $\hat{R}_{r,l}=\hat{V}_{r,l}\hat{D}_{r,l}\hat{W}_{r,l}$ with unitary operators $\hat{W}_{r,l}$ and $\hat{V}_{r,l}$ and a diagonal matrix $\hat{D}_{r,l}=\sum_{i=0}^{d-1}\lambda^{r,l}_i\ket{i} \bra{i}$. From the completeness relation $\sum_l \hat{R}_{r,l}^{\dag} \hat{R}_{r,l} = \hat{\openone}$, the singular values for the reversing operators satisfy 
\begin{equation}
\label{eq:comp2nd}
\sum_{l}\sum_{i} (\lambda^{r,l}_i)^2 =d.
\end{equation}
Without loss of generality, we can set $\hat{V}_{r,l}=\hat{\openone}$ so that the completeness relation can be written by $\sum_{l}(\lambda^{r,l}_i)^2 =1$.
\\

\subsection{Information Gain}
\label{asec:IG}

For a given ${\bf M}=\{\hat{M}_r\}$, the fidelity between the input and the estimated states $|\langle \psi \ket{\widetilde{\psi}_r}|^2$ can be averaged over all possible input states and the measurement outcomes $r$ as
\begin{equation}
\label{aeq:efidelity} {\cal \bar{G}}({\bf M}) = \int d \psi \sum_{r=1}^{n} p(r,\psi) |\langle \widetilde{\psi}_r | \psi \rangle|^2.
\end{equation}
Let us introduce the Schur's lemma, which can be used for any operator $\hat{O}$ in $d\times d$ Hilbert space;
\begin{align}
\label{aeq:schur} \int_{G} &dg \Big(\hat{U}^{\dag}_g\otimes\hat{U}^{\dag}_g \Big) \hat{O}
\left[\hat{U}_g\otimes\hat{U}_g \right] = \alpha_{1}
\hat{\openone}\otimes\hat{\openone} + \alpha_{2} \hat{S}, \\
\nonumber
&\alpha_{1}=\frac{d^2\mathrm{Tr}[\hat{O}]-d\mathrm{Tr}[\hat{O}\hat{S}]}{d^2(d^2-1)}
,~\alpha_{2}=\frac{d^2\mathrm{Tr}[\hat{O}\hat{S}]-d\mathrm{Tr}[\hat{O}]}{d^2(d^2-1)},
\end{align}
where $\hat{U}_g$ is a unitary representation of $d$-dimensional unitary group $g \in G = \mbox{U}(d)$ such that $\int_{G} dg = 1$ and $\hat{S}$ is a swap operator defined as $\hat{S} |i\rangle \otimes \ket{j} =|j\rangle \otimes \ket{i}$. 
Then, Eq.~\eqref{aeq:efidelity} is written as
\begin{equation}
\nonumber
\sum_{r=1}^{n} \int d  \psi \bra{\psi} \otimes \bra{\psi} \Big(\hat{M}_r^\dagger \hat{M}_r \otimes \ket{\widetilde{\psi}_r}\bra{\widetilde{\psi}_r}\Big)\ket{\psi} \otimes \ket{\psi},
\end{equation}
which can be recast by the Schur's lemma into
\begin{equation}
{\cal \bar{G}}({\bf M})=\frac{1}{d(d+1)}\bigg[d+\sum_{r=1}^{n} \bra{\widetilde{\psi}_r}\hat{M}_r^\dagger \hat{M}_r\ket{\widetilde{\psi}_r}\bigg].
\end{equation}
The second term can be written again as $\bra{\widetilde{\psi}_r}\hat{M}_r^\dagger \hat{M}_r\ket{\widetilde{\psi}_r}=\bra{\widetilde{\psi}_r} \hat{D}^\dag_r \hat{V}^\dag_r \hat{V}_r\hat{D}_r\ket{\widetilde{\psi}_r}=\sum_i \bra{\widetilde{\psi}_r} \hat{D}^\dag_r \ket{i}\bra{i} \hat{D}_r\ket{\widetilde{\psi}_r}=\sum_i(\lambda^r_i)^2|\bracket{i}{\widetilde{\psi}_r}|^2$. Then, the {\em information gain} can be defined as its maximum value, obtained when the estimate state is $\ket{\widetilde{\psi}_r}=\ket{0}$ for outcome $r$ so that
\begin{equation}
\label{aeq:Gmax} {\cal G}\equiv{\cal \bar{G}}_{\max}({\bf M}) =\frac{1}{d(d+1)}\bigg[d+\sum_{r=1}^n(\lambda_0^r)^2\bigg].
\end{equation}
Note that this is valid for arbitrary input states $\rho$, since the maximum evaluated in the space of pure states is also the maximum evaluated in the space of mixed states due to the convex structure of quantum states.

\subsection{Disturbance}
\label{asec:MaxF}

The fidelity between the input and the output states $|\bracket{\psi_r}{\psi}|^2$, for a given quantum measurement ${\bf M}=\{\hat{M}_r\}$, can be averaged over all possible input states and measurement outcomes $r$ as
\begin{equation}
\label{aeq:ofidelity} {\cal \bar{F}}({\bf M}) = \int d \psi \sum_{r=1}^{n} p(r,\psi) |\langle \psi_r | \psi \rangle|^2=\int d\psi \sum^n_{r=1} P(r,\psi)\frac{\bra{\psi}\hat{M}_r\ket{\psi}\bra{\psi}\hat{M}^\dag_r\ket{\psi} }{P(r,\psi)}=\int d\psi \sum^n_{r=1} \Big| \bra{\psi}\hat{M}_r\ket{\psi}\Big|^2.
\end{equation}
By the Schur's lemma, we can rewrite it as
\begin{equation}
\label{aeq:of2}
{\cal \bar{F}}({\bf M}) = \frac{1}{d(d+1)}\bigg[d+\sum_{r=1}^{n} \big| \mathrm{Tr} \hat{M}_r\big|^2\bigg].
\end{equation}
Since $| \mathrm{Tr} \hat{M}_r|=|\sum_i\bra{i}\hat{V}_r\hat{D}_r\ket{i}|=|\sum_i\bra{i}\hat{V}_r\hat{D}_r\ket{i}|=|\sum_i\lambda_i^r\bra{i}\hat{V}_r\ket{i}|\leq\sum_i\lambda_i^r|\bra{i}\hat{V}_r\ket{i}|\leq \sum_i\lambda_i^r$,
the maximum operation fidelity is then given as
\begin{equation}
\label{aeq:maxopfid} {\cal F}\equiv{\cal \bar{F}}_{max}({\bf M})=\frac{1}{d(d+1)}\bigg[d+\sum_{r=1}^n \Big(\sum_{i=0}^{d-1}\lambda_i^r\Big)^2\bigg],
\end{equation}
from which the {\em disturbance} can also be defined as the minimum operation infidelity ${\cal D}=1-{\cal F}$.

\subsection{Maximum Operation Fidelity after Reversal}
\label{asec:MaxgRev}

For given ${\bf M}=\{\hat{M}_r\}$ and ${\bf R}=\{\hat{R}_{r,l}\}$, the operation fidelity including all output states without postselection is 
\begin{equation}
{\cal \bar{F}}({\bf R}\circ{\bf M})=\int d\psi \sum^n_{r=1} \sum^m_{l=1}P(r,l,\psi) |\langle \psi_{r,l} | \psi \rangle|^2,
\end{equation}
where the right hand side can be written as
\begin{equation}
\label{eq:fidrev}
\int d\psi \sum^n_{r=1} \sum^m_{l=1}P(r,l,\psi)\frac{\bra{\psi}\hat{R}_{r,l}\hat{M}_r\ket{\psi}\bra{\psi}\hat{M}^\dag_r\hat{R}^\dag_{r,l} \ket{\psi} }{P(r,l,\psi)}=\int d\psi \sum^n_{r=1} \sum^m_{l=1}\Big| \bra{\psi}\hat{R}_{r,l}\hat{M}_r\ket{\psi}\Big|^2.
\end{equation}
From Eq.~\eqref{aeq:of2}, it can be written again as
\begin{equation}
{\cal \bar{F}}({\bf R}\circ{\bf M})=\frac{1}{d(d+1)}\bigg[d+ \sum^n_{r=1} \sum^m_{l=1}\Big| \mathrm{Tr} \hat{R}_{r,l}\hat{M}_r\Big|^2\bigg].
\end{equation}

Since 
\begin{equation}
\Big|\mathrm{Tr} \hat{R}_{r,l}\hat{M}_r\Big|=\Big|\sum_i \bra{i} \hat{D}_{r,l}\hat{W}_{r,l} \hat{V}_r\hat{D}_r \ket{i}\Big|=
\Big|\sum_i \lambda_i^r \lambda_i^{r,l} \bra{i} \hat{W}_{r,l} \hat{V}_r \ket{i}\Big|\leq\sum_i \lambda_i^r \lambda_i^{r,l} \Big|\bra{i} \hat{W}_{r,l} \hat{V}_r \ket{i}\Big|,
\end{equation}
the maximum of ${\cal \bar{F}}({\bf R}\circ{\bf M})$ is obtained when $\hat{W}_{r,l} \hat{V}_r=\hat{\openone}$. Therefore,
\begin{equation}
\label{aeq:gRmax}
{\cal \bar{F}}_{\max}({\bf R}\circ{\bf M})=\frac{1}{d(d+1)}\bigg[d+\sum_{r=1}^n  \sum^m_{l=1} \Big(\sum_{i=0}^{d-1}\lambda_i^r\lambda_i^{r,l}\Big)^2\bigg].
\end{equation}

\subsection{Reversibility}
\label{asec:MaxoRev}

We assume that $\hat{R}_{r,l}$ for $l=1,2,...,s$ are associated with the success reversal events such that 
\begin{equation}
\label{aeq:sucRev}
\hat{R}_{r,l}\hat{M}_r\ket{\psi}=\eta_{r,l}\ket{\psi},
\end{equation}
with a complex variable $\eta_{r,l}$. Since $\hat{\openone} - \sum^{s}_{l=1}\hat{R}^\dag_{r,l}\hat{R}_{r,l}$ is positive definite from the completeness relation, 
\begin{equation}
\label{aeq:upper} \sup_{\ket{\phi}}\bra{\phi}\sum^{s}_{l=1}\hat{R}^\dag_{r,l}\hat{R}_{r,l}\ket{\phi}\leq1
\end{equation}
is satisfied for arbitrary quantum state $\ket{\phi}$. Then,
\begin{equation}
\label{eq:lowersup} 
\sup_{\ket{\phi}}\bra{\phi}\sum^{s}_{l=1}\hat{R}^\dag_{r,l}\hat{R}_{r,l}\ket{\phi}\geq
\sup_{\ket{\psi_r}}\bra{\psi_r}\sum^{s}_{l=1}\hat{R}^\dag_{r,l}\hat{R}_{r,l}\ket{\psi_r}=\sup_{\ket{\psi}}\frac{\bra{\psi}\hat{M}^{\dag}_r\sum^{s}_{l=1}\hat{R}^\dag_{r,l}\hat{R}_{r,l}\hat{M}_r\ket{\psi}}{p(r,\psi)}=\frac{\sum^s_{l=1}|\eta_{r,l}|^2}{\inf_{\ket{\psi}}p(r,\psi)},
\end{equation}
so that $\sum^s_{l=1}|\eta_{r,l}|^2\leq\inf_{\ket{\psi}}p(r,\psi)$ is satisfied. For an arbitrary input state $\ket{\psi}=\sum_{i=0}^{d-1}\alpha_i\ket{i}$ where $\sum_{i=0}^{d-1}|\alpha_i|^2=1$, $\inf_{\ket{\psi}}p(r,\psi)$ is obtained when $\alpha_{d-1}=1$ and all other $\alpha_i$ are zero, because the singular values of $\hat{M}_r$ are assumed to be defined in decreasing order, so that 
\begin{equation}
\label{aeq:limitv}
\sum^s_{l=1}|\eta_{r,l}|^2\leq\inf_{\ket{\psi}}p(r,\ket{\psi})=(\lambda^r_{d-1})^2.
\end{equation}
Therefore, the {\em reversibility} can be defined as its maximum
\begin{equation}
\label{aeq:revmax}
{\cal R}\equiv\max_{\{\hat{R}_{r,l}\}} \int d\psi \sum^n_{r=1}\sum^s_{l=1}| \bra{\psi}\hat{R}_{r,l}\hat{M}_r\ket{\psi}|^2=\sum_{r=1}^{n}(\lambda^r_{d-1})^2.
\end{equation}

%%%%%%%%%%%%%%%%%%%%%%%%%%%%%%%%%%%%%%%
%\begin{figure}[b]
%\centering
%\includegraphics[width=3.4in]{figureSa}
%\caption{Left: Comparison the upper bounds by $\cal G$-$\cal D$-$\cal R$ and $\cal G$-$\cal D$ relations with the measurement in Sec.~\ref{asec:gdrex}. The bound by $\cal G$-$\cal D$-$\cal R$ is tighter than the one by $\cal G$-$\cal D$. Right: The gap is due to the reversibility $\cal R$, which is also in trade-off relation with the information gain $\cal G$.
%}\label{fig:figa}
%\end{figure}
%%%%%%%%%%%%%%%%%%%%%%%%%%%%%%%%%%%%%%%%%%

\section{Proof of Lemma 1}
\label{asec:Lemma1}

For an optimally chosen ${\bf R}=\{\hat{R}_{r,l}\}$ to reverse ${\bf M}=\{\hat{M}_r\}$, the singular values are given by $\lambda_i^{r,l=1}=\lambda_{d-1}^{r}/\lambda_{i}^{r}$. If we define $\vec{u}_i^{l}=(\lambda_{i}^{r=1}\lambda_{i}^{r=1,l},\ldots, \lambda_{i}^{r=n}\lambda_{i}^{r=n,l})$, the second term of ${\cal F}({\bf R}\circ{\bf M})$ in Eq.~\eqref{eq:maxgenrev}
\begin{equation}
\sum_{r=1}^n  \sum^m_{l=1}\Big (\sum_{i=0}^{d-1}\lambda_i^r\lambda_i^{r,l}\Big)^2=\sum_{i,j}\sum_l \vec{u}_i^{l}\ldots\vec{u}_j^{l}=\sum_{i,j}(\vec{u}_i^{l=1}\cdot\vec{u}_j^{l=1}+\sum_{l\neq1} \vec{u}_i^{l}\cdot\vec{u}_j^{l})=d^2\sum_r(\lambda^r_{d-1})^2+\sum_{i,j}\sum_{l\neq1} \vec{u}_i^{l}\cdot\vec{u}_j^{l},
\end{equation}
and $\sum_{i,j}\sum_{l\neq1} \vec{u}_i^{l}\cdot\vec{u}_j^{l}\geq\sum_{i}\sum_{l\neq1} |\vec{u}_i^{l}|^2$, where the equality is reached with a condition $\vec{u}_i^{l\neq1}\cdot\vec{u}_j^{l\neq1}=\delta_{i,j}|\vec{u}_i^{l}|^2$. From the completeness relation of the overall measurements,
\begin{equation}
\begin{aligned}
\nonumber
\sum_{r=1}^n \sum^m_{l=1}\sum_{i=0}^{d-1}(\lambda_i^r\lambda_i^{r,l})^2=d\sum_{r}(\lambda^r_{d-1})^2+\sum_{i}\sum_{l\neq1} |\vec{u}_i^{l}|^2=d.
\end{aligned}
\end{equation}
Therefore, 
\begin{align}
\nonumber
{\cal F}({\bf R}\circ{\bf M})&=\frac{1}{d(d+1)}\bigg[d+d^2\sum_{r}(\lambda^r_{d-1})^2+\sum_{i,j}\sum_{l\neq1} \vec{u}_i^{l}\cdot\vec{u}_j^{l}\bigg]\\
&\geq\frac{1}{d(d+1)}\big[2d+(d^2-d){\cal R}\big],
\end{align}
where ${\cal R}=\sum_{r}(\lambda^r_{d-1})^2$. As a result, we obtain
\begin{equation}
2+(d-1){\cal R} \leq (d+1){\cal F}({\bf R}\circ{\bf M}),
\end{equation}
with the equality condition $\vec{u}_i^{l\neq1}\cdot\vec{u}_j^{l\neq1}=\delta_{i,j}|\vec{u}_i^{l}|^2$.
$\square$\\

\section{Proof of Theorem 1}
\label{asec:Theorem1}

Let us first define $\vec{v}_i=(\lambda^{r=1}_i, \ldots, \lambda^{r=n}_i)$. Then, $g\equiv d(d+1){\cal G}-d=\sum_r(\lambda_0^r)^2=|\vec{v}_0|^2$,  $f\equiv d(d+1){\cal F}-d=\sum_r(\sum_i\lambda_i^r)^2=\sum_{i,j}\vec{v}_i\cdot\vec{v}_j$, and ${\cal R}=\sum_r(\lambda^r_{d-1})^2=|\vec{v}_{d-1}|^2$. The completeness relation in Eq.~\eqref{eq:comp1st} can be written as $\sum_r\sum_i(\lambda^r_i)^2=\sum_i|\vec{v}_i|^2=d$.

From the Schwarz inequality, 
\begin{equation}
f\leq \sum^{d-1}_{i,j=0}|\vec{v}_i||\vec{v}_j|=\Big(\sum_i|\vec{v}_i|\Big)^2=\bigg(\sqrt{g}+\sqrt{{\cal R}}+\sum^{d-2}_{i=1}|\vec{v}_i|\bigg)^2,
\end{equation}
where the equality can be reached when all the vectors $\vec{v}_i$ are collinear. Then, from the inequality of arithmetic and quadratic means,
\begin{equation}
\sum^{d-2}_{i=1}|\vec{v}_i| \leq \sqrt{(d-2)\sum^{d-2}_{i=1}|\vec{v}_i|^2},
\end{equation}
where the equality can be reached when $|\vec{v}_1|=\cdots=|\vec{v}_{d-2}|$. Here, the right hand side can be rewritten by the completeness relation as $\sqrt{(d-2)(d-g-{\cal R})}$. Therefore, we obtain
\begin{equation}
\sqrt{f}\leq \sqrt{g}+\sqrt{{\cal R}}+\sqrt{(d-2)(d-g-{\cal R})},
\end{equation}
and equivalently
\begin{equation}
\sqrt{{\cal F}-\frac{1}{d+1}}\leq \sqrt{{\cal G}-\frac{1}{d+1}} + \sqrt{\frac{{\cal R}}{d(d+1)}}+\sqrt{(d-2)\bigg(\frac{2}{d+1}-{\cal G}-\frac{{\cal R}}{d(d+1)}\bigg)}.
\end{equation}
$\square$\\

\section{Proof that $\cal G$-$\cal D$-$\cal R$ tightens $\cal G$-$\cal D$}
\label{asec:GDvsGDR}

Assume that we have $d-1$ non-negative real numbers $x_i$ where $i=1,\cdots,d-1$. From the inequality between arithmetic and quadratic mean,
\begin{equation}
\bigg(\frac{1}{d-1}\sum^{d-1}_{i=1}x_i\bigg)^2\leq\frac{1}{d-1}\sum^{d-1}_{i=1}x^2_i.
\end{equation}
where the equality holds if and only if $x_1=x_2=\cdots=x_{d-1}$. By letting $x_1=\sqrt{{\cal R}}$ and $x_2=\cdots=x_{d-1}=\sqrt{(d-g-{\cal R})/(d-2)}$ with $g\equiv d(d+1){\cal G}-d$, we obtain
\begin{equation}
\sqrt{{\cal R}}+\sqrt{(d-2)(d-g-{\cal R})}\leq\sqrt{(d-1)(d-g)},
\end{equation}
which indicates that $\cal G$-$\cal D$-$\cal R$ is tighter than $\cal G$-$\cal D$. $\square$\\

\section{Proof of Lemma 2}
\label{asec:l2}

If we define a vector $\vec{w}_i^{r}=(\lambda_{i}^{r}\lambda_{i}^{r,l=1},\ldots, \lambda_{i}^{r}\lambda_{i}^{r,l=m})$, the second term of ${\cal F}({\bf R}\circ{\bf M})$ in Eq.~\eqref{eq:maxgenrev} can be written by
\begin{equation}
\sum_{r=1}^n  \sum^m_{l=1} \bigg(\sum_{i=0}^{d-1}\lambda_i^r\lambda_i^{r,l}\bigg)^2=\sum_{r=1}^n  \sum_{i,j} \vec{w}_i^r \cdot\vec{w}_j^r.
\end{equation}
Then, by the Schwarz inequality, 
\begin{equation}
\sum_{r=1}^n  \sum_{i,j} \vec{w}_i^r \cdot\vec{w}_j^r \leq \sum_{r=1}^n  \sum_{i,j} \big|\vec{w}_i^r \big| \big|\vec{w}_j^r\big|= \sum_{r=1}^n  \Big(\sum_{i}\big|\vec{w}_i^r \big|\Big)^2=\sum_{r=1}^n  \bigg(\sum_{i}\lambda^r_i\sqrt{\sum_l(\lambda^{r,l}_i)^2}\bigg)^2,
\end{equation}
where the equality holds if all the vectors $\vec{w}_i^{r}$ are collinear. From the completeness relation for the reversing operators in Eq.~\eqref{eq:comp2nd}, we arrive at
\begin{equation}
\sum_{r=1}^n  \sum^m_{l=1} \bigg(\sum_{i=0}^{d-1}\lambda_i^r\lambda_i^{r,l}\bigg)^2\leq \sum_{r=1}^n \bigg(\sum_{i=0}^{d-1}\lambda_i^r\bigg)^2,
\end{equation}
and, from Eq.~\eqref{eq:maxopfid}, 
\begin{equation}
{\cal F}({\bf R}\circ{\bf M})\leq {\cal F}({\bf M}).
\end{equation}
$\square$\\

\section{Reversibility with errors}
\label{asec:EE}

Using a non-optimal quantum measurement or operation, the reversing process may not be able to exactly reverse the quantum measurement due to errors so that the output state is not exactly the same as the input state, s.t. $({\bf R}\circ{\bf M})(\rho) \propto \rho'\neq\rho$. Let us consider the case that the reversing operation succeeds after measurement as 
\begin{equation}
\hat{R}_{r,1}\hat{M}_r\ket{\psi}\bra{\psi}\hat{M}_r^\dag\hat{R}_{r,1}^\dag=|\eta_{r,1}|^2\rho(\psi,\epsilon)
\end{equation}
but the resulting output state $\rho(\psi,\epsilon)$ deviates from the original input state. Here, we consider an arbitrary error model parameterized by $0\leq \epsilon \leq 1$ and the error density function $p(\epsilon)$, s.t. $\int^1_0 p(\epsilon) d\epsilon = 1$. We assume that the error occurs independently of the measurement outcome for simplicity. 

From the definition of the reversibility in Eq.~\eqref{eq:revmax}, we can evaluate 
\begin{equation}
\begin{aligned}
{\cal R}(\epsilon) &= \max_{\{\hat{R}_{r,1}\}} \int d\psi \sum_{r} \bra{\psi}\hat{R}_{r,1}\hat{M}_r\ket{\psi}\bra{\psi}\hat{M}_r^\dag\hat{R}_{r,1}^\dag\ket{\psi}\\
&= \max_{\{\hat{R}_{r,1}\}} \int d\psi \sum_{r} |\eta_{r,1}|^2 \bra{\psi}\rho(\psi,\epsilon)\ket{\psi}\\
&= \sum_r (\lambda^r_{d-1})^2  \int d\psi \bra{\psi}\rho(\psi,\epsilon)\ket{\psi}.
\end{aligned}
\end{equation}
By averaging over the whole error region, we obtain 
\begin{equation}
{\cal R}=\int^1_0 d\epsilon~p(\epsilon) {\cal R}(\epsilon) 
= \sum_r(\lambda^r_{d-1})^2 F_{s},
\end{equation}
where $F_{s}=\int d\psi \int^1_0 d\epsilon~p(\epsilon)\bra{\psi}\rho(\psi,\epsilon)\ket{\psi}$ is defined as the average fidelity between the input state and the output state when the reversing operation succeeds.

\section{Proof of $\cal{S}_{\cal G\text{-}D\text{-}R} \cap \cal{S}_{\cal G\text{-}R}=S_{\cal G\text{-}D}$}
\label{asec:SetP}

We can prove that a quantum measurement satisfying both saturation conditions of $\cal G$-$\cal D$-$\cal R$ and $\cal G$-$\cal R$ always saturates $\cal G$-$\cal D$. From the condition to saturate $\cal G$-$\cal D$-$\cal R$ described in Eq.~\eqref{eq:scon}, the singular values satisfy $\lambda^{r}_0\geq\lambda^{r}_1=\cdots=\lambda^{r}_{d-2}\geq\lambda^{r}_{d-1}$. The measurement operator is then written by $\hat{M}_r=\lambda^{r}_0\ket{0}\bra{0}+\lambda^{r}_1(\ket{1}\bra{1}+\cdots+\ket{d-2}\bra{d-2})+\lambda^{r}_{d-1}\ket{d-1}\bra{d-1}$ so that
\begin{equation}
\label{eq:pfinter}
\hat{M}^{\dag}_r\hat{M}_r=\big\{(\lambda^{r}_0)^2-(\lambda^{r}_{d-1})^2\big\}\ket{0}\bra{0}+\big\{(\lambda^{r}_1)^2-(\lambda^{r}_{d-1})^2\big\}\big(\ket{1}\bra{1}+\cdots+\ket{d-2}\bra{d-2}\big)+(\lambda^{r}_{d-1})^2\hat{\openone}.
\end{equation}
If it satisfies the saturation condition of $\cal G$-$\cal R$, i.e., $\hat{M}^{\dag}_r\hat{M}_r=a_r\ket{j_r}\bra{j_r}+b_r\hat{\openone}$ with non-negative $a_r$ and $b_r$ where $j_r \in \{0,\ldots,d-1\}$ \cite{Cheong12}, we find that
\begin{equation}
\lambda^{r}_1=\lambda^{r}_{d-1}.
\end{equation} 
As a result, the quantum measurement satisfies the saturation condition of $\cal G$-$\cal D$, i.e., all $\vec{v}_i$ are collinear and $|\vec{v}_1|= \cdots =|\vec{v}_{d-1}|$, where $\vec{v}_i=(\lambda^{r=1}_i, \ldots , \lambda^{r=n}_i)$ \cite{Banaszek01}. $\square$\\

\end{document}